\documentclass[12pt]{article} 

\usepackage{latexsym} 
\usepackage{amssymb}  
\usepackage{amsbsy}   
\usepackage{graphicx}       

\newcommand{\al}{\alpha}
\newcommand{\be}{\beta}

\newcommand{\Ga}{\Gamma}
\newcommand{\de}{\delta}
\newcommand{\De}{\Delta}

\newcommand{\eps}{\epsilon}

\newcommand{\La}{\Lambda}

\newcommand{\<}{\langle} 
\renewcommand{\>}{\rangle} 

\newcommand{\txt}{\textstyle}

\newcommand{\dsp}{\displaystyle}

\newcommand{\ad}{\dagger}
\newcommand\eqn[1]{(\ref{#1})}      
\newcommand\Eqn[1]{Eq.~(\ref{#1})}  
\newcommand{\e}{{\rm e}}
\newcommand{\beq}{\begin{equation}}
\newcommand{\eeq}{\end{equation}}
\newcommand{\ba}{\begin{array}}
\newcommand{\bea}{\begin{eqnarray}}
\newcommand{\ea}{\end{array}}
\newcommand{\eea}{\end{eqnarray}}

\newcommand\comment[1]{ \hbox{[{\it Comment suppressed here.}\/]} }
\newcommand\hide[1]{}

\newcommand{\tr}{\hbox{tr}}

\newcommand{\skipover}[1]{}

\newcommand{\half} {{\txt \frac{1}{2}}}
\newcommand{\third}{{\txt \frac{1}{3}}}

\makeatletter 

\def\appendix{\par                              
    \setcounter{section}{0}                     
    \setcounter{subsection}{0}
    \renewcommand{\theequation}{\Alph{section}.\arabic{equation}}
    \renewcommand{\thesection}{Appendix \Alph{section}
                \setcounter{equation}{0}  } 
}

\def\applabel#1{\@bsphack
  \protected@write\@auxout{}%
         {\string\newlabel{#1}{{\Alph{section}}{\thepage}}}%
  \@esphack}

\def\section{
\setcounter{equation}{0}        
\@startsection {section}{1}{\z@}{-3.5ex plus -1ex minus 
 -.2ex}{2.3ex plus .2ex}{\large\bf}}
\renewcommand{\theequation}{\arabic{section}.\arabic{equation}}

\def\subsection{\@startsection{subsection}{2}{\z@}{-3.25ex plus -1ex minus 
 -.2ex}{1.5ex plus .2ex}{\normalsize\bf}}

\def\subsubsection{\@startsection{subsubsection}{3}{\z@}{-3.25ex plus
 -1ex minus -.2ex}{1.5ex plus .2ex}{\normalsize}}

\makeatother   

\newsavebox{\eqlabel}

\makeatletter  
\newlength{\numblen}
\newsavebox{\eqnumb}
\def\@eqnnum{\savebox{\eqnumb}{\rm (\theequation)}%
\settowidth{\numblen}{\usebox{\eqnumb}}%
\makebox[\numblen][l]{\usebox{\eqnumb}~~~\usebox{\eqlabel}}}
\makeatother   

\newenvironment{equationwithlabel}[1]{ %
  \begin{equation}\label{#1} }{\end{equation}} 
\newcommand{\beql}[1]{\begin{equationwithlabel}{#1}}
\newcommand{\eeql}{\end{equationwithlabel}}

\newcommand{\MeV}{{\rm MeV}} 
\newcommand{\GeV}{{\rm GeV}} 
\newcommand{\vp}{{\mathbf p}}    
\newcommand{\vk}{{\mathbf k}}    
\newcommand{\vq}{{\mathbf q}}    
\newcommand{\vr}{{\mathbf r}}    
\newcommand{\dm}{{\delta\mu}} 
\newcommand{\mubar}{{\bar\mu}}
\newcommand{\dmmax}{{\de\mu_2}}
\newcommand{\dmmin}{{\de\mu_1}}

\begin{document}

\title{{\bf Crystalline Color Superconductivity}}



\author{
	Mark Alford, Jeffrey A. Bowers, and Krishna Rajagopal \\[0.5ex]
{\normalsize Center for Theoretical Physics}\\
{\normalsize Massachusetts Institute of Technology}\\
{\normalsize Cambridge, MA 02139 }
}

\newcommand{\preprintno}{
  \normalsize MIT-CTP-3012 
}

\date{August 18, 2000 \\[1ex] \preprintno}

\begin{titlepage}
\maketitle
\def\thepage{}          

\begin{abstract}
In any context in which color superconductivity arises
in nature, it is likely to involve pairing between species of quarks
with differing chemical potentials.   For suitable
values of the differences between chemical potentials,
Cooper pairs with nonzero total momentum 
are favored, as was first realized by Larkin, Ovchinnikov,
Fulde and Ferrell (LOFF).   Condensates of this sort spontaneously
break translational and rotational invariance, leading to
gaps which vary periodically in a crystalline pattern.
Unlike the original LOFF state, these crystalline
quark matter 
condensates include both spin zero and spin one Cooper pairs.
We explore
the range of parameters for which crystalline
color superconductivity arises in the QCD phase diagram.
If in some shell within the quark matter core
of a neutron star (or within a strange quark star)  
the quark number densities are
such that crystalline color superconductivity arises,
rotational vortices may be pinned in this shell, making
it a locus for glitch phenomena.
\end{abstract}

\end{titlepage}

\renewcommand{\thepage}{\arabic{page}}

\section{Overview}
\label{sec:overview}

The attraction between two quarks which are anti-symmetric in color
renders cold dense quark matter unstable to the formation
of quark Cooper pairs in a color superconducting
state~\cite{Barrois,BailinLove,ARW2,RappEtc,BergesRajagopal,ARW3,Reviews}.
If two (or more) different quark flavors are involved,
and their Fermi momenta are the
same, they pair as in the standard BCS state. The pairing is guaranteed
because in the absence of an interaction each pair costs no free
energy---each quark can be created at its Fermi surface---and the
interaction then makes the system unstable against formation of a
condensate of pairs.

In this paper we study the situation, generic in the real world,
where the Fermi momenta of the two species are different. 
If the Fermi momenta are far apart, no pairing between the species is
possible. The transition between the BCS and unpaired states as the
splitting between Fermi momenta increases has been studied in
electron~\cite{Clogston} and QCD~\cite{ABR2+1,SW2,Bedaque}
superconductors, assuming that no other state intervenes.  However,
there is good reason to think that another state can occur.  This is
the ``LOFF'' state, first explored by Larkin and Ovchinnikov~\cite{LO}
and Fulde and Ferrell~\cite{FF} in the context of electron
superconductivity in the presence of magnetic impurities.
They found that near the
unpairing transition, it is favorable to form a crystalline state in
which the Cooper pairs have nonzero momentum. This is favored because
it gives rise to a region of phase space where each of the two quarks
in a pair can be close to its Fermi surface,
and such pairs can be created at low cost in free energy.

We study the pairing between two species whose chemical potentials
differ by $2\dm$ and find that
for a large class of interactions there is a window of $\dm$ within
which states of the LOFF type are preferred over the BCS and unpaired states.
This has important ramifications for compact star
phenomenology, since it means that there may be a layer of crystalline
quark matter inside the star. This could pin rotational vortices,
and lead to the kind of glitch phenomena that have up to now
been thought of as uniquely associated with the nuclear
crust of neutron stars.

In Section~\ref{sec:intro}, we give a more detailed introduction to
the BCS and LOFF color superconducting states, and their possible
astrophysical applications.  
In Section~\ref{sec:LOFF}, we describe
the LOFF state in quark matter with $\dm\neq 0$.  We note in
particular that, unlike in the original LOFF context, there is pairing
both in $J=0$ and $J=1$ channels.  
In Section~\ref{sec:gap}, we
derive the gap equation for the LOFF state for a model Hamiltonian in
which the full QCD interaction is replaced by a 
four-fermion interaction with the quantum numbers of 
single gluon exchange.
In Section~\ref{sec:results}, we use the gap
equation to evaluate the range of $\dm$ within which the LOFF state
arises. We will
see that at low $\dm$ the translationally invariant BCS state, with
gap $\De_0$, is favored. At $\dm_1$ there is a first order transition
to the LOFF paired state, which breaks translational symmetry. At
$\dm_2$ all pairing disappears, and translational symmetry is
restored at a phase transition which is
second order in mean field theory.  
In the weak-coupling limit, in which $\Delta_0\ll \mu$, we
find values of $\dm_1$ and $\dm_2$ which are in quantitative agreement
with those obtained by LOFF. This agreement occurs only because we 
have chosen
an interaction which is neither attractive nor repulsive in the
$J=1$ channel, making the $J=1$ component of our LOFF condensate
irrelevant in the gap equation.  
In Section~\ref{sec:ham}, we consider
a more general Hamiltonian in which the 
couplings corresponding to electric and magnetic
gluon exchange can be separately tuned. 
This leads to interactions in both $J=0$
and $J=1$ channels, and we show how it affects
the range of $\dm$ within which the LOFF state arises.  
In Section~\ref{sec:conclusions}, we outline
future work which follows immediately from what we have done and look
farther ahead toward possible astrophysical consequences of
crystalline color superconductivity.

We recommend that the astrophysically inclined reader, interested
primarily in the consequences of our results, read Sections
\ref{sec:intro} and \ref{sec:conclusions}, skipping those in between.

\section{Introduction}
\label{sec:intro}

\subsection{Astrophysical applications of color superconductivity}

Our current understanding of the color superconducting
state of quark matter leads us to believe that it
may occur naturally in compact stars. 
The critical temperature below which quark matter 
is a color superconductor is generally estimated to be
of order $10-50~\MeV$, which suggests that 
any quark matter which occurs within
neutron stars that are more than a few seconds old
is in a color superconducting state.

This estimate of the critical temperature
comes both from models whose parameters are tuned to
reproduce zero density 
physics~\cite{ARW2,RappEtc,BergesRajagopal,ARW3,CD,RSSV2,EHS,SW0} 
and also from weak coupling
methods which are quantitatively valid at asymptotically high
densities~\cite{Son,SW3,PR,Hong,HMSW,BLR,HS,Schaefer,BBS,ShovWij,EHHS,RajagopalShuster}, 
with chemical potentials $\mu \gg 10^8~\MeV$~\cite{RajagopalShuster}. 
Neither class of methods can be trusted
quantitatively for quark number chemical potentials $\mu\sim 400$ MeV,
as appropriate for the quark matter which may occur in the cores of
neutron stars.  Still, both methods agree that the gaps at
the Fermi surface are of order tens to 100 MeV, with critical temperatures
about half as large.

It is therefore important to look for astrophysical consequences of
color superconductivity.
As a Fermi surface phenomenon, it has
little effect on the equation of state, and hence little effect
on the radius of a compact star.
There are nevertheless several effects of color superconductivity
under active investigation.  The
color superconductivity of quark matter in neutron stars influences
the evolution of magnetic fields within the quark matter~\cite{ABRmag}
(see also~\cite{BlaschkeMag}).  Cooling by neutrino emission is also
affected~\cite{Prakash} (see also~\cite{BlaschkeCool}).  In quark
stars, the physics of the instability to r-mode oscillations is
dramatically affected by color superconductivity~\cite{Madsen}, although
this is not the case for neutron stars with quark matter present only
in their cores~\cite{BildstenUshomirsky,Madsen}. 
Furthermore, the phase transition at which color
superconductivity sets in as a hot proto-neutron star cools may yield
a detectable signature in the neutrinos received from a
supernova~\cite{CarterReddy}.  Finally,
one goal of the present paper is to motivate an
investigation of the possibility that (some) pulsar glitches may
originate in quark matter.

If two species of fermion experience an attractive interaction, and
their Fermi momenta are the
same, they pair in the standard BCS state. The pairing is guaranteed
because in the absence of an interaction each pair costs no free
energy (each quark can be created at its Fermi surface), and the
interaction then makes the system unstable against formation of a
condensate of pairs.  In the QCD context, if there are two flavors
of quarks with equal Fermi momenta, quarks of two colors 
and two
flavors pair~\cite{ARW2,RappEtc} while if there are three flavors of quarks,
all nine quarks pair in a pattern which locks color and flavor
symmetries, breaking chiral symmetry~\cite{ARW3,SWcont}.  These idealizations
are very instructive, but in any physical context, the up, down and
strange quarks will all have different Fermi momenta.
To give the reader some sense for typical scales in
the problem, we give an illustrative example~\cite{ABRmag}.
The numbers in this 
paragraph assume that the
quarks are noninteracting fermions---clearly a bad
assumption---and so should certainly not be construed 
as precise.
Consider quark matter with average
quark chemical potential $\mu=400~\MeV$, made of
massless up and down quarks and strange quarks
with mass $M_s=300~\MeV$.  ($M_s$ is a density
dependent effective mass; this adds to the uncertainty
in its value.)  If the strange quark were massless,
quark matter consisting of equal parts $u$, $d$ and $s$
would be electrically neutral.  
In our illustrative example,
on the other hand, electric neutrality requires a nonzero
density of electrons, with chemical potential $\mu_e=53~\MeV$.
Charge neutrality combined with the requirement that the
weak interactions are in equilibrium determine all the
chemical potentials and Fermi momenta:
\beql{int:illustrative}
\ba{rcrcl@{\qquad}rcl}
\mu_u &=& \mu - \frac{2}{3} \mu_e &=& 365~\MeV,
   & p_F^u  &=&  \mu_u, \\[2ex]
\mu_d &=& \mu + \frac{1}{3} \mu_e &=& 418~\MeV,
   & p_F^d  &=&  \mu_d, \\[2ex]
\mu_s &=& \mu + \frac{1}{3} \mu_e &=& 418~\MeV, 
   & p_F^s &=& \sqrt{\mu_s^2-M_s^2} = 290~\MeV, \\[2ex]
 && \mu_e &=& 53~\MeV,  & p_F^e &=& \mu_e~.
\ea
\label{fiducial}
\eeql
The baryon number density 
$\rho_B = (1/3\pi^2) [(p_F^u)^3 +(p_F^d)^3 + (p_F^s)^3]$ 
is 4 times nuclear matter density.\footnote{Had we chosen 
$M_s=200$ MeV, we would have obtained $\mu_e=24$ MeV,
$p_F^u=384$ MeV, $p_F^d=408$ MeV, $p_F^s=356$ MeV and a $\rho_B$
of 5 times nuclear matter density.}  As one goes deeper into
a neutron star, $\mu$ increases, $M_s$ decreases
somewhat, and $\mu_e$ and all differences between 
the quark Fermi momenta decrease.\footnote{Note that in 
a neutron star with a quark matter core, regions of
purely hadronic and purely quark matter are separated
by a mixed phase in which neither
the hadronic regions nor the quark matter regions are
separately charge neutral~\cite{Glendenning}.
The electrically charged quark matter in these regions will
have Fermi momenta which differ qualitatively from those
in our example, with $p_F^u$ 
less than either $p_F^d$ or $p_F^s$~\cite{Glendenning}.}
In this paper, we investigate the consequences
of pairing between quarks with differing Fermi momenta.
For simplicity, we restrict our explicit calculations to
the case of two massless quarks with differing chemical potentials
$\mu_u$ and $\mu_d$, which we write as   
\beql{int:mu}
\mu_d = \bar\mu + \dm;\quad \mu_u = \bar\mu - \dm.
\eeql
We expect similar phenomena to those
we describe to arise wherever any one of $|p_F^u - p_F^d|$  or 
$|p_F^u - p_F^s|$ or $|p_F^d - p_F^s|$ falls within a 
suitable range, but we leave the investigation of quark
matter with $u$, $d$, and massive $s$ quarks to future work.
We also work at zero temperature throughout.

\subsection{Isotropic (non-LOFF) pairing}

In the color superconducting phase for two massless quark
flavors at the same chemical potential $\mu$,
the condensate consists of quark-quark pairs which are
flavor singlets and color $\mathbf{\bar 3}$ antitriplets (and hence
also spin singlets, to obey Pauli statistics).  Pairing is of the BCS
type: a red up quark of momentum $\vp$ pairs with a green down quark of
momentum $-\vp$ of the same helicity, so that the spins are
antiparallel.  The blue quarks are left unpaired.
Such pairing is strongest
in the vicinity of the Fermi surface, for 
$||\vp|-\mu| \lesssim \Delta_0$, where $\Delta_0$ is the BCS gap
parameter. 

If, instead, 
the Fermi momenta are sufficiently different, no BCS pairing is
possible.  It is no longer possible to guarantee that the formation of
pairs lowers the free energy, because in the BCS state the two
fermions in a pair have equal and opposite momentum, so at most one
member of each pair can be created at its Fermi surface. The other
member costs non-zero free energy, which the attractive interaction may be
unable to compensate.

Assuming that no other state intervenes between the BCS state
and the state with no condensate, we can
apply the results first derived by Clogston and 
Chandrasekhar
in the context of pairing between spin-up and spin-down electrons
with differing Fermi momenta~\cite{Clogston}.  For small enough 
$\delta \mu$, the favored BCS state has coincident Fermi
surfaces, 
$p_F^u = p_F^d = \bar\mu$ because this maximizes the pairing
and thus the gain
in interaction energy.\footnote{If one tries to
construct a ``BCS-like'' state which has
$p_F^u=\mu_u$ and $p_F^d=\mu_d$ and
consequently no pairing for $p_F^u<p<p_F^d$,
one finds~\cite{Sarma}
that this state has a
higher free energy than the BCS state (in which  
$p_F^u = p_F^d = \bar\mu$).   The gain in free energy
associated with choosing $p_F^u=\mu_u$ and $p_F^d=\mu_d$
does not compensate for the lost pairing energy.}
We denote the gap in this BCS state by $\Delta_0$.
The free energy of this BCS state must be compared
to that of the
unpaired or ``normal'' state in which 
the quarks simply distribute themselves in Fermi
seas with $p_F^u = \mu_u$, $p_F^d = \mu_d$ and no condensate forms. 
The BCS state is the stable ground state of the
system only when its negative interaction energy offsets the large
positive free energy cost associated with forcing the Fermi seas to
deviate from their normal state distributions.  
If $\Delta_0\ll \bar\mu$ and $\dm\ll \bar\mu$, the free
energy of the BCS state relative to that of the normal state
at a given $\dm$ is
\beql{int:FBCS}
F_{\rm BCS} - F_{\rm normal} = \frac{\bar\mu^2}{\pi^2}\left(
2\dm^2-\Delta_0^2\right)\ .
\eeql
The coefficient $1/3\pi^2$ depends on the number of fermion species
which pair and is appropriate to the case of interest to us.
Clogston and Chandrasekhar concluded
that the BCS state is favored for $\dm<\dm_1=\Delta_0/\sqrt{2}$. 
(The relation $\dm_1=\Delta_0/\sqrt{2}$ is exact only
in the weak-coupling limit in which $\Delta_0 \ll \bar\mu$.)  At 
$\dm=\dm_1$, there is a first order phase transition at which
the gap parameter drops discontinuously from $\Delta_0$ to zero:
for $\dm < \dm_1$, the system is in the BCS phase, unperturbed from the
the $\dm = 0$ state.

This analysis is modified in an interesting way
at nonzero temperature, as was discussed by Lombardo and Sedrakian in
the context of pairing between neutrons and protons in
nuclei~\cite{Sedrakian}.  Thermal excitations smear out the normal state 
Fermi surfaces, making
pairing between thermally excited states above the lower Fermi surface
and below the upper Fermi surface possible.
As a consequence, as $T$ is increased from zero,
there is a range of $T$ within which $\dm_1$ 
is larger than at $T=0$.
At still higher temperatures, of course, all 
pairing is lost.

In applying the work of Clogston and
Chandrasekhar to color superconductivity, there have
been two extensions to their analysis.
First, recall that only two colors of up and
down quarks pair.  In describing the mixed phase associated
with the first order phase transition, one must take careful
account of the unpaired blue quarks.
This has been done by Bedaque~\cite{Bedaque}.
Second, in Refs.~\cite{ABR2+1,SW2} the transition between
the color-flavor locked phase and the two-flavor color
superconducting phase has been studied, under the
assumption that $\mu_u=\mu_d=\mu_s$ but with $p_F^s\neq p_F^{u,d}$
because of the nonzero strange quark mass.  
The first order transition that these authors describe
is similar to that of Clogston and Chandrasekhar, as
it is associated with the unpairing of $us$ and $ud$ Cooper
pairs, but it differs in that the analogue of the
normal state is one in which $u$ and $d$ quarks remain paired.
As we have seen above, treating a realistic situation
requires relaxing the assumption of equal chemical potentials.

\subsection{Non-isotropic (LOFF) pairing}

The Clogston and Chandrasekhar analysis of the first
order unpairing transition assumes that the only possible
phases of the system are a BCS phase and the normal phase.
However, there is good reason to think that another state can occur in
the crossover region between BCS and no pairing.  As was first
realized by Larkin and Ovchinnikov~\cite{LO} and Fulde and
Ferrell~\cite{FF} (LOFF), whereas the BCS state requires pairing
between fermions with equal and opposite momenta, when $\dm\sim\dm_1$
it may be more favorable to form a condensate of Cooper pairs with
{\it nonzero} total momentum.  By pairing quarks with momenta which
are not equal and opposite, some Cooper pairs are allowed to have both
the up and the down quarks on their respective Fermi surfaces even
when $\dm\neq 0$.  LOFF found that within a range of $\dm$ near
$\dm_1$, a condensate of Cooper pairs with momenta $\vq+\vp$ and
$\vq-\vp$ (see Figure~\ref{fig:angles}) is favored over either the BCS
condensate or the normal state. Here, our notation is such that $\vp$
specifies a particular Cooper pair, while $\vq$ is a fixed vector,
the same for all pairs,
which characterizes a given LOFF state.  The magnitude $|\vq|$ is
determined by minimizing the free energy; the direction of $\vq$ is
chosen spontaneously.  The resulting LOFF state breaks translational
and rotational invariance.  In position space, it describes a
condensate which varies as a plane wave with wave vector $2\vq$.

Once one has demonstrated an instability to the formation of a plane
wave, it is natural to expect that the state which actually develops
has a crystalline structure. Larkin and Ovchinnikov in fact argue that
the favored configuration is a crystalline condensate which varies in
space like a one-dimensional standing wave, $\cos(2\vq\cdot\vr)$. Such
a condensate vanishes along nodal planes~\cite{LO}.  Subsequent
analyses suggest that the crystal structure may be more complicated.
Shimahara~\cite{Shimahara} has shown that in two dimensions, the LOFF
state favors different crystal structures at different temperatures: a
hexagonal crystal at low temperatures, square at higher temperatures,
then a triangular crystal and finally a one-dimensional standing wave
as Larkin and Ovchinnikov suggested at temperatures that are higher
still. In three dimensions, the question of which
crystal structure is favored seems unresolved~\cite{Buzdin}.

LOFF did their analysis in the same context as that
of Clogston and Chandrasekhar: electromagnetic superconductivity
in a magnetic field which causes a Zeeman splitting while not
inducing screening currents.  They were seeking to model
the physics of magnetic impurities in a superconductor.
Magnetic effects on the motion of the electrons~\cite{Gruenberg}
and the scattering of electrons off non-magnetic 
impurities~\cite{Aslamazov,Takada2} disfavor the LOFF 
state. Although signs of the BCS to LOFF transition
in the heavy fermion superconductor UPd$_2$Al$_3$
have been reported~\cite{Gloos}, the interpretation
of these experiments is not unambiguous~\cite{Controversy}.
It has also been suggested that the LOFF phase
may be more easily realized in condensed matter systems
which are two-dimensional~\cite{Shimahara,2D} or 
one-dimensional~\cite{1D}, both because in these cases
$\dm_2$ is larger than in three-dimensional systems and because
the magnetic field applied precisely parallel to a 
one- or two-dimensional system does not affect the motion
of electrons therein.  Evidence for a LOFF phase in
a quasi-two-dimensional layered organic 
superconductor has recently been reported~\cite{Nam}.

None of the difficulties which have beset attempts to 
realize the LOFF phase in a system of electrons in a magnetic
field arise in the QCD context of interest to us.
Differences between quark chemical potentials are generic
and the physics which leads to these differences has
nothing to do with the motion of the quarks.  We therefore
expect the original analysis of LOFF (without
the later complications added in order to  
treat the difficulties in the condensed matter physics
context) to be a good starting point.  In this paper
we use an analysis based on that originally done by
Fulde and Ferrell~\cite{FF}, 
but described in more detail by Takada and
Izuyama~\cite{Takada1},
to argue that for
appropriate values of quark number densities, the color
superconducting gap may vary periodically in space, forming
a crystalline pattern.  More precisely, what we will demonstrate
is that if some difference between chemical
potentials falls in the appropriate range, quark matter is unstable to the
spontaneous breaking of
translational invariance by the formation of condensates
which vary in space like a plane wave.  
Following Larkin and Ovchinnikov~\cite{LO}, we expect that
once there is an instability to the formation of plane waves
the condensate that results will be crystalline, but we
leave the determination
of the crystal structure of the condensate 
to future work.

\section{The LOFF state}
\label{sec:LOFF}

\begin{figure}
\begin{center}
\includegraphics[width=3.0in]{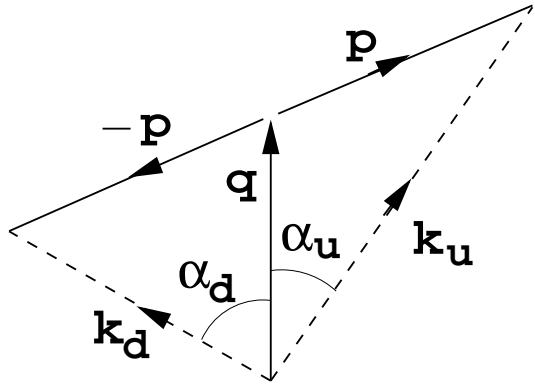}
\end{center}
\caption{
The momenta $\vk_u$ and $\vk_d$
of the two members of a LOFF-state Cooper pair.  We choose
the vector $\vq$, common to all Cooper pairs,
 to coincide with the $z$-axis.  The angles $\alpha_u(\vp)$ and 
$\alpha_d(\vp)$ indicate the polar angles of $\vk_u$ and $\vk_d$,
respectively.
}
\label{fig:angles}
\end{figure}

We begin our analysis of a LOFF state for quark matter by
constructing a variational ansatz for the LOFF wavefunction.  As
motivated by the preceding discussion, we consider Cooper pairs which
consist of an up quark and a down quark with respective momenta
\beql{LOFF:mom}
\vk_u = \vq+\vp,\hspace{0.3in} \vk_d =  \vq-\vp, 
\eeql
so that $\vp$ identifies a particular quark pair, and every quark pair
in the condensate has the same nonzero total momentum $2\vq$.  This
arrangement is shown in Figure~\ref{fig:angles}.  The helicity and
color structure are obtained by analogy with the ``2SC'' state as
described in previous work~\cite{ARW2,RappEtc}: the quark pairs will be color
$\mathbf{\bar 3}$ antitriplets, and in our ansatz we consider only
pairing between quarks of the same helicity.

With this in mind, here is a suitable trial wavefunction for the
LOFF state with wavevector $\vq$~\cite{LO,FF,Takada1}:
\beql{LOFF:ansatz}
\ba{rcl}
|\Psi_{\vq}\> &=& B^\ad_L B^\ad_R |0\>,\\[1ex]
B^\ad_L &=& 
  \dsp\prod_{\vp \in {\cal P}, \alpha, \beta} 
    \left( \cos \theta_L(\vp) 
    + \epsilon^{\alpha\beta3} \e^{i \xi_L(\vp)} 
    \sin \theta_L(\vp)\, a^\ad_{Lu\alpha}(\vq\!+\!\vp) 
    \, a^\ad_{Ld\beta}(\vq\!-\!\vp) \right) 
   \\[4ex]
 & \times & 
  \dsp\prod_{\vp \in {\cal B}_u, \alpha} a^\ad_{Lu\alpha}(\vq\!+\!\vp) 
  \times 
  \dsp\prod_{\vp \in {\cal B}_d, \beta} a^\ad_{Ld\beta}(\vq\!-\!\vp) 
  , \\[3ex]
B^\ad_R &=& \mbox{as above,~} L \to R,
\ea
\eeql
where $\alpha$, $\beta$ are color indices, $u$, $d$ and $L$, $R$ are
the usual flavor and helicity labels, and $a^\ad$ 
is the particle creation operator 
(for example, $a^\ad_{Ld\al}$ creates a left-handed down quark with color
$\al$).  The $\theta$'s and $\xi$'s are
the variational parameters of our ansatz: they are 
to be chosen to minimize
the free energy of the LOFF state, as described in the next section.  The
first product in equation \eqn{LOFF:ansatz} creates quark pairs within a
restricted region ${\cal P}$ of the total phase space. This allowed
``pairing region'' will be discussed below. 
The next product
fills a ``blocking region'' ${\cal B}_u$ with unpaired up quarks:
these are up quarks with momenta $\vq+\vp$ for which there are no
corresponding down quarks with momenta $\vq-\vp$.  The final
product fills the blocking region ${\cal B}_d$ with unpaired down
quarks.
The ansatz does not contain a term that would create antiparticle
pairs: we have checked the effect of such a term and found that
it has no qualitative effect on our results.
\begin{figure}[t]
\begin{center}
\includegraphics[width=5.5in]{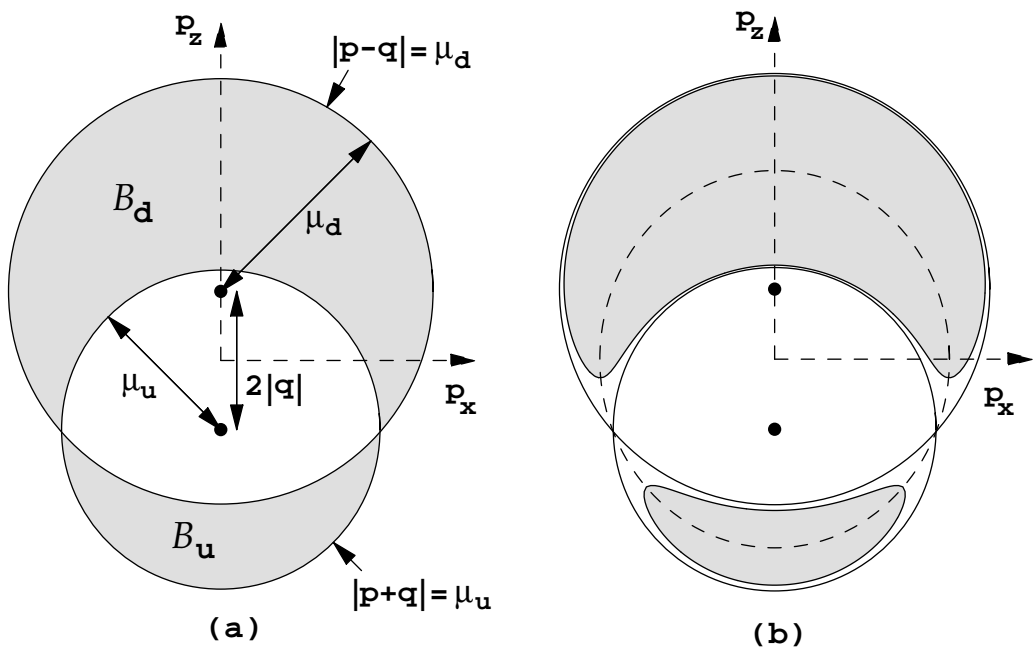}
\end{center}
\caption{
The LOFF phase space, as a function of $\vp$
(\Eqn{LOFF:mom}). We show the $p_y=0$ plane.
(a) The phase space in
the limit of arbitrarily weak interactions. In the shaded
blocking regions ${\cal B}_u$
and ${\cal B}_d$, no pairing is possible. In the inner unshaded region,
an interaction can induce hole-hole pairs. In the outer unshaded
region, an interaction can induce particle-particle pairs.
The region ${\cal P}$ (Eq.~\eqn{LOFF:ansatz})
is the whole unshaded area.
(b)~When the effects of interactions and the formation of the LOFF state
are taken into account, the blocking regions shrink.
The BCS singularity occurs on the dashed ellipse,
defined by $\eps_u+\eps_d=\mu_u+\mu_d$, where making a Cooper pair
costs no free energy in the free case.
}
\label{fig:regions}
\end{figure}

To complete the specification of 
our ansatz we need to describe the allowed pairing and
blocking regions in phase space.  These regions are largely determined
by Pauli blocking as a result of populated Fermi seas.  In the absence
of pairing interactions, the system is in the ``normal'' state and up
and down quarks are distributed in Fermi seas with Fermi momenta
$p_F^u = \mu_u$ and $p_F^d = \mu_d$, respectively (recall that we
consider massless quarks only, so the single particle energy of a
quark with momentum $\vk$ is $\epsilon(\vk) = |\vk|$).  An up quark
carries momentum $\vk_u=\vp + \vq$; in $\vp$-space, therefore, the Fermi sea
of up quarks corresponds to a sphere of radius $\mu_u = \mubar-\dm$
centered at $-\vq$.  Similarly, a down quark carries momentum 
$\vk_d=-\vp +\vq$, giving a 
sphere in $\vp$-space of radius $\mu_d = \mubar+\dm$
centered at $+\vq$.  The two offset spheres are shown in
Figure~\ref{fig:regions}a (we have drawn the 
case $|\vq| > \dm$ so that
the two Fermi surfaces intersect in $\vp$-space).  In the limit of
arbitrarily weak interactions,
the blocking region ${\cal B}_u$ corresponds to the lower shaded area
in the figure: pairing does not occur here since the region is inside
the Fermi sea of up quarks, but outside the Fermi sea of down quarks.
Similarly the upper shaded area is the blocking region ${\cal B}_d$.
The entire unshaded area is the pairing region ${\cal P}$: it includes
the region inside both spheres, where hole-hole pairing can occur, and
the region outside both spheres, where particle-particle pairing can
occur.  

We can now explain how the LOFF wavefunction ansatz can describe
the normal state with no condensate: we choose $\theta_L(\vp) =
\theta_R(\vp) = \pi/2$ for $\vp$ inside both Fermi spheres, and
otherwise all the $\theta$'s are zero.  
With this
choice the first term in \Eqn{LOFF:ansatz} fills
that part of 
each Fermi sea corresponding to the inner unshaded region of 
Figure~\ref{fig:regions}a.  The ${\cal B}_u$ and ${\cal B}_d$ terms fill
out the remainder of each Fermi sea to obtain the normal state.
Note that in the absence of pairing, the normal state can be
described with any choice of $\vq$.  The most convenient choice
is $\vq=0$, in which case $\vk_u=\vk_d=\vp$, 
${\cal B}_u$ vanishes, and ${\cal B}_d$ is a spherical shell.
Other choices of $\vq$ correspond to choosing different
origins of  $\vk_u$-space and $\vk_d$-space, but in the 
absence of any interactions this has no consequence.
Once we turn on interactions and allow pairing, we expect
a particular $|\vq|$ to be favored.

The phase space picture changes slightly when pairing interactions are
included: the blocking regions are smaller when a LOFF condensate is
present, as indicated in Figure~\ref{fig:regions}b.  We will account
for this effect in the next section.  With smaller blocking regions, a
larger portion of the phase space becomes available for LOFF pairing.
Such pairing is guaranteed to be
energetically favorable when it costs zero free 
energy to create an up quark and a down quark, since these quarks 
can then pair to obtain a negative interaction energy.  The zero free
energy condition is 
\beql{LOFF:sing}
\epsilon(\vk_u)+\epsilon(\vk_d) = \mu_u + \mu_d = 2 \bar\mu
\eeql
where $\epsilon(\vk)$ is the single particle energy of a quark with
momentum $\vk$.  For massless quarks, we obtain
$|\vq+\vp|+|\vq-\vp| = 2 \bar\mu$, which describes an ellipsoidal
surface in $\vp$-space.  This surface is indicated by the ellipse
shown in Figure~\ref{fig:regions}b; notice that the ellipsoid and
the two Fermi surfaces all intersect at a circle.

If the interaction is weak, we expect LOFF pairing to be
favored in a thin layer of phase space around this ellipsoid.  This is
manifest in the gap equation derived in the next section
(Eq.~(\ref{gap:LOFF})) in which, as in BCS theory, we find a divergent
integrand on this ellipsoid in the absence of pairing.  Pairing
smoothes the divergence.  As the interaction gets stronger, the layer
of favored pairing gets thicker.  If there were no blocking regions,
we could use the entire ellipsoid, just as BCS pairs condense over the
entire spherical surface $|\vp| = \mu$ in the symmetric, $\dm=|\vq|=0$
case.  However, as shown in Figure~\ref{fig:regions}b, the blocking
regions exclude pairing over most of the ellipsoid, leaving a ribbon
of unsuppressed LOFF pairing in the vicinity of the circle where the
Fermi surfaces intersect.  This agrees with our expectation for the
particle distribution in the LOFF state: it is as in the normal state,
except that there is a restricted region (around the aforementioned ribbon)
where each quark in a pair can be near its Fermi surface.

Although the constant single-particle energy 
contours for noninteracting up and down
quarks cross in $\vp$-space (see Figure~\ref{fig:regions}a), we
emphasize that the Fermi surfaces of up and down quarks 
do not cross in momentum ($\vk_u$- and $\vk_d$-)
space.  The $\vp$-space ribbon of unsuppressed pairing
corresponds to unsuppressed pairing between up and down quarks 
with momenta around $\vk$-space ribbons near their 
respective (disjoint) Fermi surfaces.

In the limit of arbitrarily weak interactions, the ribbon in
momentum space along which pairing is unsuppressed shrinks, as
the blocking regions grow to exclude all of the ellipsoid
except the one-dimensional circle at which the two spheres in 
Figure~\ref{fig:regions} intersect.  This circle has
insufficient phase space to lead to a singularity
in the gap equation: the integrand is singular on this circle,
but the integral does not diverge.  Therefore, the LOFF
state is not guaranteed to occur if one takes the weak coupling
limit at fixed $\dm$. In this respect, the LOFF state
is like the BCS state at nonzero $\dm$: for weak coupling,
$\Delta_0\rightarrow 0$ and because the BCS state can only
exist if it has $\Delta_0>\sqrt{2}\dm$, it must vanish
for couplings weaker than some threshold.  
We shall see, however, that at any fixed weak coupling, the LOFF
state, like the BCS state, is guaranteed to occur at some $\dm$: the
BCS state arises if $\dm<\dm_1$ and the LOFF state arises if
$\dm_1<\dm<\dm_2$.

One of the most striking features of the LOFF state is the spin
structure of the condensate.  The familiar ``2SC'' state pairs quarks
of the same helicity and opposite momentum, so the spins are
antiparallel and the quarks are arranged in an antisymmetric
combination to form spin singlet Cooper pairs.  The LOFF state also
pairs quarks of the same helicity, but now the quark momenta are no
longer antiparallel, as can be seen from Figure~\ref{fig:angles}.
Therefore the LOFF Cooper pairs cannot be spin singlets: they are 
superpositions of both spin zero and spin one.  This is revealed
explicitly by evaluating the nonzero $\< \psi\psi \>$ expectation
values in the LOFF state:
\beql{LOFF:cond}
\ba{rcl}
-\< \Psi_{\vq} | \epsilon_{ij} \epsilon_{\alpha \beta 3} 
  \,\psi^{i\alpha}(\vr) C L \,\psi^{j\beta}(\vr) 
  | \Psi_{\vq} \> &=& 2\Gamma_A^L \e^{i 2\vq\cdot\vr} \\[1ex] 
i\< \Psi_{\vq} | (\sigma_1)_{ij} \epsilon_{\alpha \beta 3} 
  \,\psi^{i\alpha}(\vr) C L \sigma^{03} \,\psi^{j\beta}(\vr) 
  | \Psi_{\vq} \> &=& 2\Gamma_B^L \e^{i 2\vq\cdot\vr}  
\ea
\eeql
where $i$, $j$ are flavor indices (1 = up, 2 = down), $\alpha$,
$\beta$ are color indices, $C = i \gamma^0 \gamma^2$, $L =
(1-\gamma_5)/2$ is the usual left-handed projection operator, and
$\sigma_{\mu\nu} = (i/2)[\gamma_\mu,\gamma_\nu]$.  The constants
$\Gamma_A^L$ and $\Gamma_B^L$ are left-handed $J=0$ and $J=1$ 
condensates, respectively.  $\Gamma_A^R$ and $\Gamma_B^R$ 
are defined analogously. The
$\Gamma$'s can be expressed in terms of the variational parameters 
of the LOFF wavefunction:
\beql{LOFF:gammas}
\ba{rcl}
\Gamma_A^L & = & \dsp\frac{4}{V} \dsp\sum_{\vp \in {\cal P}} 
  \sin \theta_L(\vp) \cos \theta_L(\vp) \e^{i\xi_L(\vp)} \,
  \sin \Bigl(\frac{\al_u(\vp)+\al_d(\vp)}{2}\Bigr)\e^{-i\phi(\vp)} \\[1ex]
\Gamma_B^L & = &  \dsp\frac{4}{V} \dsp\sum_{\vp \in {\cal P}} 
  \sin \theta_L(\vp) \cos \theta_L(\vp) \e^{i\xi_L(\vp)} \,
  \sin \Bigl(\frac{\al_u(\vp)-\al_d(\vp)}{2}\Bigr)\e^{-i\phi(\vp)}
\ea
\eeql
Here $V$ is the spatial
volume of the system, $\alpha(\vp)$ are the
polar angles of the quark momenta, as in Figure~\ref{fig:angles},
and the dependence on the azimuthal angle $\phi$ follows from
our use of the spinor conventions described in 
Refs.~~\cite{BailinLove,ARW2,ARW3}.  
The expressions for $\Gamma_A^R$ and $\Gamma_B^R$  are the same
as those in (\ref{LOFF:gammas}) except
that $\phi(\vp)$ is replaced by $\pi-\phi(\vp)$.  
In Eq.~(\ref{LOFF:gammas})
and throughout, $(1/V)\sum_\vp$ becomes $\int d^3p/(2\pi)^3$
in an infinite system.

Once we have
derived a gap equation by minimizing the free energy
with respect to these variational parameters, we expect the condensates to 
be simply related to gap parameters occurring in the gap equation.
We will see explicitly how $\Gamma_A$ and $\Gamma_B$ are determined
in the next section.  

Notice that the condensates of \Eqn{LOFF:cond} are plane
waves in position space by virtue of the nonzero momentum $2\vq$ of a
Cooper pair.  $\Gamma_A$ describes pairing which is antisymmetric in
color, spin, and flavor, while $\Gamma_B$ describes pairing which is
antisymmetric in color but symmetric in spin and flavor (in each case,
Pauli statistics are obeyed).  
In the original LOFF condensate of electrons
there can be no $\Ga_B$, since electrons have no color or flavor,
so that only the spin antisymmetric pairing is possible.

The $J=0$ condensates $\< \psi C L \psi\>$, 
$\< \psi C R \psi \>$ are Lorentz scalars (mixed under parity), while
the $J=1$ condensates $\< \psi C L \sigma^{03} \psi \>$, 
$\< \psi C R \sigma^{03} \psi \>$ are 
{\bf 3}-vectors (also mixed under parity) which point in
the $z$-direction, parallel to
the spontaneously chosen direction $\hat\vq$ of the
LOFF state.
Because the ansatz contains a $J=1$
component, it would be interesting to generalize it
to include the possibility of $LR$ pairing, in addition to
$LL$ and $RR$ pairing. We discuss this further in Section~\ref{sec:ham}.

The possibility of a LOFF phase in QCD
has been mentioned briefly in a different context.
In their analysis of quark matter with a very large
isospin density (with large Fermi momenta for down and {\it anti}-up
quarks) Son and Stephanov have noted that if the $d$ and $\bar u$
Fermi momenta differ suitably, a 
LOFF phase will arise~\cite{SonStephanov}.

In the physically realizable context of large baryon number
density, pairing between quarks and holes with nonzero
total momentum
has also been 
discussed~\cite{DGR,ShusterSon,ParkRhoWirzbaZahed,RappCrystal}.
This results in a condensate with the quantum
numbers of $\langle \bar q q \rangle$, 
which varies
in space with a wave number equal to $2\mu$; in contrast, the LOFF
phase describes a diquark condensate which varies with
a wave number $2|\vq|$ comparable to $2\dm$.  The crystalline
chiral condensate~\cite{RappCrystal} is favored 
in QCD at asymptotically high densities only if the number
of colors is very large~\cite{DGR}, greater than 
about $N_c=1000$~\cite{ShusterSon,ParkRhoWirzbaZahed}.
At lower densities, where the interaction is stronger, the crystalline
chiral condensate may arise in QCD with fewer colors~\cite{RappCrystal}.
Apparently, however, in QCD with $N_c = 3$ this phase is not 
favored (although it is close to being competitive) even
when the coupling is so large that $\Delta_0/\mu > 1/2$. 
Note that crystalline color superconductivity
is guaranteed to
occur at arbitrarily weak coupling for suitably chosen $\dm$,
while a crystalline chiral condensate cannot form anywhere
in the phase diagram if the coupling is weak.

\section{The gap equation and free energy}
\label{sec:gap}

Having presented a trial wavefunction for the LOFF state, we now
proceed 
to minimize the expectation value of the free energy
$\< F \>$ with respect to the variational parameters
of the wavefunction (the $\theta$'s and $\xi$'s of equation
(\ref{LOFF:ansatz})) to obtain a LOFF gap equation.  The free energy is
$F = H - \mu_u N_u - \mu_d N_d$, where $H$ is the Hamiltonian, and
$N_u$ and $N_d$ are the number operators for up and down quarks,
respectively.  We choose
a model Hamiltonian which has a free quark term $H_0$ and an
interaction term $H_I$, and write the free energy as $F = F_0 +
H_I$, where $F_0 = H_0 - \mu_u N_u - \mu_d N_d$ is the free energy for
noninteracting quarks.  To describe the pairing interaction between
quarks, we use an NJL model consisting of a four-fermion interaction
with the color and flavor structure of one-gluon exchange:
\beql{gap:ham}
H_I = \frac{3}{8} \int d^3 x \left[ G_E (\bar\psi \gamma^0 T^A \psi) 
(\bar\psi \gamma^0 T^A \psi) -  G_M (\bar\psi \gamma^i T^A \psi)  
(\bar\psi \gamma^i T^A \psi) \right]
\eeql
where the $T^A$ are the color $SU(3)$ generators,
normalized so that $\tr(T^AT^B)=2\de^{AB}$.  
Notice that we have
relaxed some constraints on the spin structure of one-gluon exchange:
we allow for the possibility of independent couplings $G_E$ and $G_M$
for electric and magnetic gluons, respectively.  This spoils Lorentz
boost invariance but there is no reason to insist on boost invariance
in a finite-density system.  Indeed, in high density quark matter we
expect screening of electric gluons but only Landau damping of
magnetic gluons, and we might choose to model these effects by setting
$G_E \ll G_M$.  We postpone a discussion of these issues and their
implications for the LOFF state until Section~\ref{sec:ham}.  For now,
we restrict ourselves to the case of Lorentz invariant single gluon
exchange, by letting $G_E = G_M = G > 0$.

We need to
evaluate the expectation value of $F$ in the LOFF state to obtain 
an expression for the free energy of the system in terms of the variational
parameters of the ansatz.  The noninteracting part of the free energy
is simply
\beql{gap:F0}
\ba{rcl}
\< F_0 \> &=&  \dsp\sum_{\vp \in {\cal B}_u} 2 (|\vq + \vp | - \mu_u) + \sum_{\vp \in {\cal B}_d} 2 (|\vq - \vp | - \mu_d) \\
 & & + \dsp\sum_{\vp \in {\cal P}} 2 (|\vq+\vp| + |\vq-\vp| - \mu_u - \mu_d) \sin^2 \theta_L(\vp) \\
 & & + (\mbox{same, with } L \to R).
\ea
\eeql
The first and second terms represent the contributions of the unpaired
left-handed up and down quarks, respectively.  The third term gives
the (noninteracting) free energy of the left-handed quark pairs.  The
three terms are all repeated with $L$ replaced by $R$ to include the
free energy for the right-handed quarks.  The factors of two in
equation (\ref{gap:F0}) appear because there are two quark colors
(``red'' and ``green'') involved in the the condensate.  The ``blue''
quarks do not participate in the pairing interaction and instead
behave as free particles: the blue up and down quarks fill Fermi seas
with Fermi momenta $p_F^u = \mu_u$ and $p_F^d = \mu_d$,
respectively.  Below, we will
want to compare the free energy of the LOFF, BCS and normal 
states.  Since at any given $\mu_u$ and $\mu_d$ 
the free energy of the spectator quarks
is the same in all three states, we 
can neglect these blue quarks in the 
remainder of our analysis even though they do contribute
to the total free energy.  


The expectation value of $H_I$ gives the total binding energy 
of the pairing interaction:
\beql{gap:HI}
\< H_I \> = -\half GV \left( |\Gamma^L_A|^2 + |\Gamma^R_A|^2 \right)
\eeql
where the $\Gamma_A$'s are the $J=0$ LOFF condensates defined
in equations (\ref{LOFF:gammas}).  
These condensates 
are simply related to $J=0$ LOFF gap parameters 
defined as 
\beql{gap:DeltaDefn}
\Delta^{\{L,R\}}_A = G \Gamma_A^{\{L,R\}}\ .
\eeql
The gap parameters $\Delta_A$ correspond to 1PI Green's functions
and are the quantities which will appear in the
quasiparticle dispersion relations 
and for which we will derive the self-consistency conditions
conventionally called gap equations.
We see from Eq.~(\ref{gap:HI}) that with $G > 0$ the interaction
is attractive in the $J=0$ channel and is neither attractive
nor repulsive in the $J=1$ channel.  

Our ansatz breaks rotational invariance, so once $J=0$ pairing
occurs ($\Ga_A\neq 0$) we expect that 
there will also be $J=1$ pairing
($\Ga_B\neq 0$). As we have seen, this arises
even in the absence of any interaction in the $J=1$ channel
as a consequence of the fact that the momenta of
two quarks in a Cooper pair are not anti-parallel if $\vq\neq 0$. 
Because $\langle H \rangle$
is independent of $\Gamma_B$, the quasiparticle 
dispersion relations must also be independent of $\Gamma_B$.  
That is, the $J=1$ gap parameter 
must vanish: $\De_B=0$.
In Section~\ref{sec:ham},
we shall see by direct calculation that
$\Delta_B$ is proportional to $(G_E-G_M)\Gamma_B$. In the
present analysis with $G_E=G_M$, therefore, $\Delta_B=0$ while
$\Gamma_B\neq 0$.

The $\xi$'s  are chosen
to cancel the azimuthal phases $\phi(\vp)$ in equations
(\ref{LOFF:gammas}). 
By this choice we obtain maximum coherence
in the sums over $\vp$, giving the largest possible magnitudes for 
the condensates and gap parameters.  We have
\beql{gap:phases}
\xi_L(\vp) = \phi(\vp) + \varphi_L, \hspace{0.3in} 
\xi_R(\vp) = \pi-\phi(\vp) + \varphi_R
\eeql
where $\varphi_L$ and $\varphi_R$ are arbitrary $\vp$-independent
angles.  
These constant phases do not affect the free 
energy --- they correspond to the Goldstone bosons for the
broken left-handed and right-handed baryon number symmetries --- and
are therefore not constrained by the variational procedure.  
For convenience, we set $\varphi_L = \varphi_R = 0$ 
and obtain condensates
and gap parameters that are purely real.  

The relative phase $\varphi_L - \varphi_R$ 
determines how the LOFF condensate transforms under a parity
transformation. Its value determines whether
the $J=0$ condensate is scalar, pseudoscalar,
or an arbitrary combination of the two and whether
the $J=1$ condensate is vector, pseudovector, or an arbitrary combination.
Because single gluon exchange cannot change the handedness of a
massless quark, the left- and right-handed 
condensates in the LOFF phase are not coupled in the 
free energy of Eq.~(\ref{gap:HI}.)
Our choice of 
interaction Hamiltonian
therefore allows an arbitrary choice of $\varphi_L - \varphi_R$.
A global $U(1)_A$ transformation changes $\varphi_L - \varphi_R$,
and indeed this is a symmetry of our toy model.
If we included $U(1)_A$-breaking interactions in
our Hamiltonian, to obtain a more complete description of QCD,
we would find that the free energy depends
on $\varphi_L - \varphi_R$, and thus selects a 
preferred value. For example, had we taken $H_I$ to be the
two-flavor instanton interaction as in Ref.~\cite{ARW2,RappEtc}, 
the interaction energy would appear as 
$\Gamma^{L*}\Gamma^R + \Gamma^L \Gamma^{R*}$ instead
of as in (\ref{gap:HI}). This would
enforce a fixed phase relation $\varphi_L-\varphi_R=0$,
favoring condensates which are parity conserving~\cite{ARW2,RappEtc}.


We now apply the variational method to determine the angles
$\theta(\vp)$ in our trial wavefunction, by requiring that the free
energy is minimized: $\partial \< F \> / \partial \theta(\vp) = 0$.
This is complicated by the fact that the pairing region ${\cal P}$ and
the blocking regions ${\cal B}_u$ and ${\cal B}_d$ are themselves
implicitly dependent on the $\theta$ angles: these angles determine
the extent of the LOFF pairing, and the phase space regions ${\cal P}$,
${\cal B}_u$ and ${\cal B}_d$
change
when a condensate is present, as mentioned in Section~\ref{sec:LOFF}.
For now we simply ignore any $\theta$-dependence of the phase space
regions; our result will nevertheless turn out to be correct.  Everything is
the same for left and right condensates so we hereafter drop the $L$
and $R$ labels.  Upon variation with respect to $\theta(\vp)$,
we obtain
\beql{gap:tan2th}
\tan 2\theta(\vp) = \frac{ 2 \Delta_A \sin (\beta_A(\vp)/2) }{|\vq+\vp| 
+ |\vq-\vp| - \mu_u - \mu_d}
\eeql
where $\beta_A(\vp) = \alpha_u(\vp) + \alpha_d(\vp)$ is the angle
between the two quark momenta in a LOFF pair, as shown in 
Figure~\ref{fig:angles}.  Notice that the denominator on the right hand side
of the above expression vanishes along the ellipsoidal surface of
optimal LOFF pairing described in Section~\ref{sec:LOFF}.  When $\vq = 0$,
the quark momenta are antiparallel so $\beta_A(\vp) = \pi$ and 
Eq.~(\ref{gap:tan2th}) reduces to the simple BCS result:
$\tan 2\theta = \Delta_A/(|\vp|-\mubar)$.  

With the $\theta$ angles now expressed in terms of a gap parameter
$\Delta_A$, we turn to the LOFF
quasiparticle dispersion relations. They can be obtained
by taking the absolute value of the expressions
\beql{gap:quasi}
\ba{rcrcl}
E_1(\vp) &=& \dm &+& \half( |\vq+\vp| - |\vq-\vp|) \\[1ex]
 && & +& \half \sqrt{ (|\vq+\vp| + |\vq-\vp| - 2\bar\mu)^2 
+ 4 \Delta_A^2 \sin^2(\half\beta_A(\vp))} \\[2ex]
E_2(\vp) &=& -\dm & -& \half( |\vq+\vp| - |\vq-\vp|) \\[1ex]
 && & +& \half \sqrt{ (|\vq+\vp| + |\vq-\vp| - 2\bar\mu)^2 
+ 4 \Delta_A^2 \sin^2(\half\beta_A(\vp))}\ , \\
\ea
\eeql
whose meaning we now describe.
For regions of $\vp$-space which are well outside both Fermi surfaces,
$E_1$ ($E_2$) is the free energy cost of removing
a LOFF pair and adding an up quark with momentum 
$\vq+\vp$ (a down quark with momentum $\vq-\vp$).  
For regions of $\vp$-space which are well inside both Fermi surfaces,
$E_1$ ($E_2$) is the free energy cost of removing
a LOFF hole pair and adding a down hole with momentum 
$\vq-\vp$ (an up hole with momentum $\vq+\vp$).  
Where the Fermi surfaces cross in $\vp$-space and pairing
is maximal, both quasiparticles are equal superpositions of
up and down.  
In the region of $\vp$-space which is well inside the up Fermi
surface but well outside the down Fermi surface, 
$E_1$ is negative, corresponding
to a domain in which it is energetically favorable to have an unpaired
up quark with momentum $\vq+\vp$ rather than a $(\vq+\vp,\vq-\vp)$
quark pair.  Similarly, $E_2$ is negative where it is favorable
to have an unpaired down quark with momentum $\vq-\vp$ rather than a LOFF pair.
Equations (\ref{gap:quasi}) allow us to finally 
complete our description of the LOFF phase by specifying the 
definitions of the phase space regions ${\cal P}$,
${\cal B}_u$ and ${\cal B}_d$.
The blocking region ${\cal B}_u$ 
is the region where $E_1(\vp)$ is negative, and unpaired up quarks
are favored over LOFF pairs.
Similarly ${\cal B}_d$ is the region where $E_2(\vp)$ is
negative. 
The regions $E_1 < 0$ and $E_2 < 0$ are shown as the shaded
areas in Figure~\ref{fig:regions}a for $\Delta_A = 0$, and in 
Figure~\ref{fig:regions}b for $\Delta_A \neq 0$.  LOFF pairing occurs in
the region where $E_1$ and $E_2$ are both positive:
\beql{gap:pregion}
{\cal P} = \{ \vp | E_1(\vp) > 0 \mbox{ and } E_2(\vp) > 0 \}
\eeql
corresponding to the entire unshaded regions of Figure~\ref{fig:regions}.
The actual quasiparticle dispersion functions are $|E_1(\vp)|$ and
$|E_2(\vp)|$: they are nonnegative everywhere, since they represent
energies of 
perturbations of the LOFF state which is the presumed ground state of
the system.\footnote{
Since the LOFF condensate contains pairs with momentum $2\vq$,
the momentum of its quasiparticle excitations is only defined
modulo $2\vq$. The momentum, modulo $2\vq$, 
of a quasiparticle of energy $|E_1(\vp)|$ is $\vp{\rm ~mod~} 2\vq$.}
In the blocking regions, elementary excitations are
created by replacing an unpaired quark with a quark pair, and vice
versa in the pairing region.
When $\vq=0$, Eqs.~(\ref{gap:quasi}) reduce to the more familiar BCS
result: $E_{\{1,2\}}(\vp) = \pm \dm + \sqrt{(|\vp|-\bar\mu)^2 + \Delta_A^2}$.

With the boundaries of the blocking regions specified, one
can verify 
by explicit calculation
that the variation of these boundaries 
upon variation of the $\theta$'s does not change the free energy.
This can be understood as follows.
Notice that 
because we can create zero-energy quasiparticles on
the boundaries of the blocking regions, 
there is no actual energy gap in the
excitation spectrum of the LOFF state.
The change in $\< F \>$ 
due to variation of the boundaries of the 
blocking regions is zero because this
variation simply creates zero-free-energy 
quasiparticles on these boundaries.
This justifies our neglect of the 
$\theta$-dependence of the phase space
regions in the derivation of Eq.~(\ref{gap:tan2th}).

Substituting the expression (\ref{gap:tan2th}) for the $\theta$ angles 
into the expression (\ref{LOFF:gammas}) for the $\Gamma_A$ condensate,
and using the relation $\Delta_A = G
\Gamma_A$, we obtain a self-consistency equation for the gap parameter
$\Delta_A$:
\beql{gap:LOFF}
1 = \frac{2G}{V}\sum_{\vp\in {\cal P}} 
\frac{2\sin^2(\half\beta_A(\vp))}{\sqrt{
  (|\vq+\vp|+|\vq-\vp|-2\bar\mu)^2 + 4\Delta_A^2\sin^2(\half\be_A(\vp))
  }
}.
\eeql
This can be compared to the BCS gap equation, obtained upon setting
$\vq=0$ and eliminating the blocking regions:
\beql{gap:BCSgap}
1 = \frac{2G}{V}\sum_\vp \frac{1}{\sqrt{(|\vp|-\bar\mu)^2+\Delta_0^2}}\ .
\eeql
Note that in the LOFF gap equation (\ref{gap:LOFF}),
the gap parameter appears on the right hand side 
both explicitly in the denominator 
and also implicitly in the definition of the pairing region
${\cal P}$, as given in (\ref{gap:pregion}).   
This means that if
the $\vq\rightarrow 0$ limit is taken at fixed $\dm$,
the LOFF gap equation will only become the BCS gap equation
if the blocking regions vanish in this limit. This happens
if, as $\vq\rightarrow 0$, $\De_A$
tends to a limiting value which is greater than $\dm$.  
A state with $\De_A<\dm$ and $\vq=0$
is ``BCS-like'', in that the Cooper
pairs have zero momentum,  but has no pairing 
within a region $p_F^u<|\vp|<p_F^d$.  Such states always
have higher free energy than the BCS state obtained simply
by solving the gap equation (\ref{gap:BCSgap}), appropriate
if there are no blocking regions and $p_F^u=p_F^d$~\cite{Sarma}.
 
In the next section we will solve the LOFF gap equation 
(\ref{gap:LOFF}) and
determine the circumstances in which the LOFF state is the true ground
state of the system.
Once we have obtained a solution to the gap equation (\ref{gap:LOFF}) for 
$\Delta_A$, the condensates
are given by $\Gamma_A = \Delta_A/G$ 
and 
\beql{gap:GammaB}
\Gamma_B = \dsp\frac{2}{V} \sum_{\vp \in {\cal P}} \frac{ 2 \Delta_A 
\sin(\half\beta_A(\vp))\sin (\half\beta_B(\vp))}{ \sqrt{ ( |\vq+\vp| +
|\vq-\vp| - 2\bar\mu)^2 + 4 \Delta_A^2 \sin^2(\half\beta_A(\vp)) }}
\eeql
where $\beta_B(\vp)=\alpha_u(\vp)-\alpha_d(\vp)$. 
(See Figure~\ref{fig:angles}.)    
We now see explicitly that if the interaction is 
attractive in the $J=0$ channel, creating a nonzero
$\Gamma_A$ and $\Delta_A$, a nonzero $J=1$
condensate $\Gamma_B$ is induced regardless of the
fact that there is no interaction in the $J=1$ channel.
As a check, note that
if $\vq=0$, $\sin(\half\beta_A(\vp))=1$ and 
$\sin(\half\beta_B(\vp))$ is given by the cosine of the polar
angle of $\vp$.  The right hand side of (\ref{gap:GammaB}) 
therefore vanishes upon integration, and $\Gamma_B$ vanishes when $\vq=0$
as it should.
It is now apparent that two features contribute to a nonzero $\Ga_B$.
The first is that the momenta in a quark pair are not antiparallel,
which leads to the factors of $\sin(\half\be_A(\vp))$ in \Eqn{gap:GammaB}.
The second is that the pairing region is anisotropic, since if it were
not the factor of $\sin(\half\be_B(\vp))$ would ensure that the
right-hand side of \eqn{gap:GammaB} vanishes upon integration.

As written, the gap equations (\ref{gap:LOFF})
and (\ref{gap:BCSgap}) are
ultraviolet divergent.  In QCD,
of course, asymptotic freedom implies that the interaction between
quarks decreases at large momentum transfer and we have
not yet represented this fact in our toy model.   
In previous work~\cite{ARW2,ARW3,BergesRajagopal}, 
we chose to mimic the effects of asymptotic
freedom (and to render the right hand side of the gap equation finite)
by introducing a form factor associated with each fermion leg
in the four-fermion interaction.  This is not a good strategy
when $\vq\neq 0$.  The two incident quarks carry
momenta $\vq+\vp$ and $\vq-\vp$ while the 
outgoing quarks carry momenta $\vq+\vp'$ and $\vq-\vp'$.
Were we to cut off these four momenta with form
factors on each leg, we would have a
cutoff which depends explicitly on $\vq$.  
This is not a good representation
of what happens in full QCD, in which the condition
for when the interaction becomes weak is determined by
the momentum $\vp-\vp'$ transferred through the gluon  and
has nothing to do with $\vq$.
For simplicity, 
we choose to introduce a hard cutoff in our NJL model, rather
than a smooth form factor, and choose simply to cut off
the momentum $\vp$.  This is not equivalent to cutting
off the momentum transfer, but has the desired feature
of being a  $\vq$-independent cutoff.
That is, we limit the integration region
to $|\vp|<\Lambda$ in the BCS gap equation (\ref{gap:BCSgap})  
and to $\{\vp \in {\cal P}$ and $|\vp|<\Lambda\}$ in
the LOFF gap equation (\ref{gap:LOFF}).  In the BCS case,
this criterion is equivalent to cutting off the momentum
of each fermion leg. In the LOFF case, it is not equivalent
and is more appropriate.  The choice we have made is not
the only $\vq$-independent cutoff one might try.
For example, 
we have also obtained results upon cutting off
momenta outside a large ellipsoid in $\vp$-space, confocal
with the centers of the two Fermi spheres in Figure~\ref{fig:regions},
but have found that this makes little difference relative
to the simpler choice of the large sphere $|\vp|<\Lambda$.


\section{Results}
\label{sec:results}
We solve the gap equation \eqn{gap:LOFF} numerically
(and analytically in the limit $\De_A \ll \dm,q,\De_0$) and
calculate the LOFF state free energy as a function of $\dm$ and $q$, for given
coupling $G$, average chemical potential $\mubar$, and cutoff $\La$.
We vary $q$ to minimize the LOFF free energy, and
compare it with that for the standard BCS
pairing \eqn{gap:BCSgap} to see which is favored.
In this way we can map out the phase diagram for the three
phases of pairing between the two species of quark:
BCS, LOFF, and unpaired.

Note that the solution to the gap equation, the LOFF gap parameter
$\De_A$, is not a gap in the spectrum of excitations.  
The quasiparticle dispersion 
relations (\ref{gap:quasi}) vary
with the direction of the momentum, yielding gaps that vary from zero
(for momenta on the edge of the blocking regions in phase space) up to
a maximum of $\De_A$.

We will first discuss the range of $\dm$ in which
there exists a LOFF state as a local energy minimum.
Later we will go on to
study the competition between LOFF and BCS, and see in what range
of $\dm$ the LOFF state
is the global minimum.
We expect the BCS state to be preferred when the mismatch $\dm$
between the Fermi energies of the two species is small.
When the mismatch is comparable to the BCS gap
($\dm\sim\De_0$) we expect a transition to LOFF, and
at larger $\dm$ we expect all pairing to cease.
These expectations are largely borne out.

In general we fix $\La=1~\GeV$ and $\mubar=0.4~\GeV$, and study
different coupling strengths $G$ which we parameterize by
the physical quantity $\De_0$, the BCS gap of Eq.~\eqn{gap:BCSgap}
which increases monotonically with increasing $G$.
When we wish to study the dependence on the cutoff, we
vary $\La$ while at the same time 
varying the coupling $G$ such that $\De_0$ is kept fixed.  
(This is in the same spirit as using a renormalization
condition on a physical quantity---$\De_0$---to fix
the ``bare'' coupling---$G$.)  We expect that the relation between
other physical quantities
and $\De_0$ will be reasonably insensitive to variation
of the cutoff $\La$.

\begin{figure}[t]
\begin{center}
\includegraphics[width=2.1in,angle=-90]{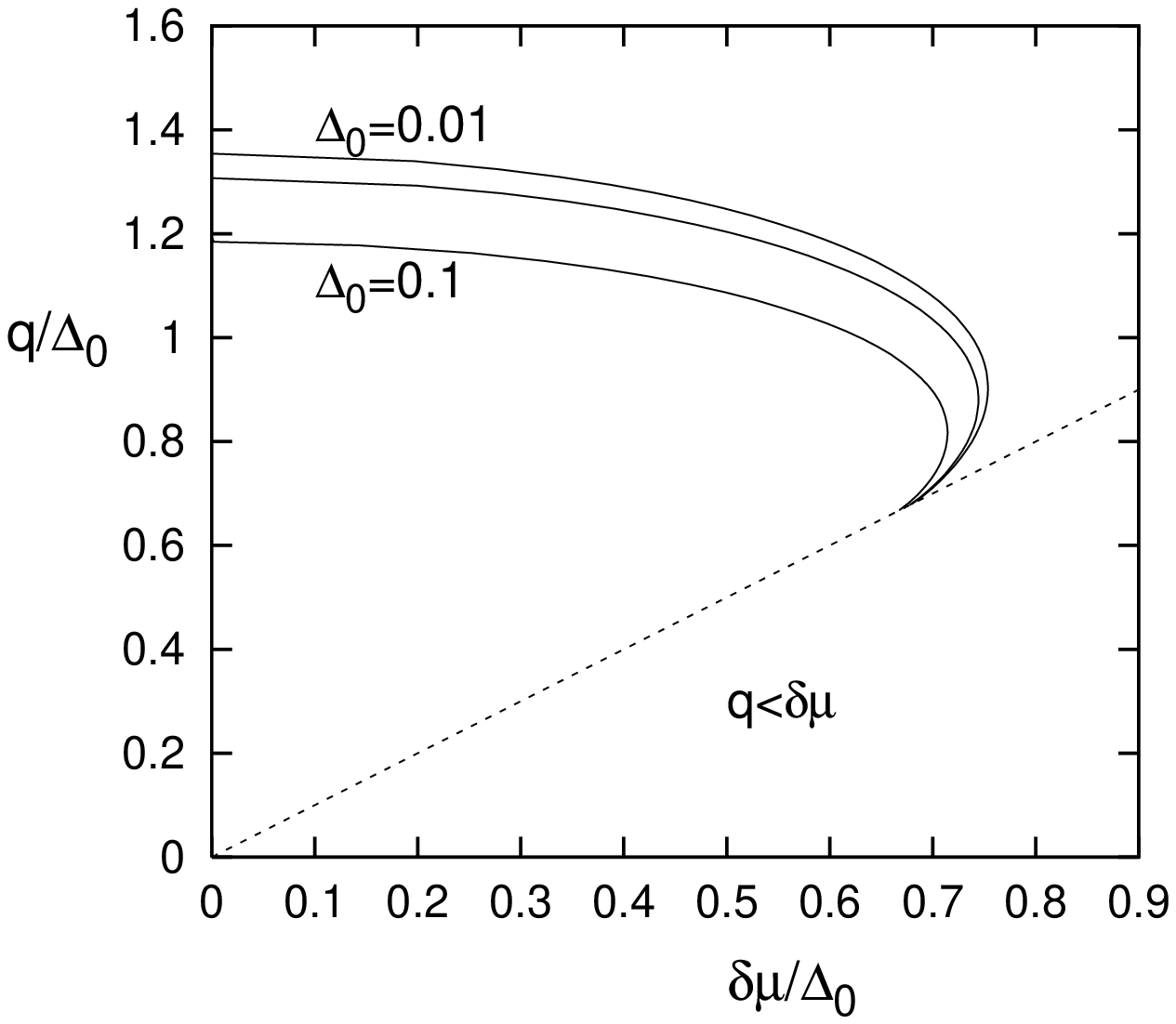}
\phantom{XX}
\includegraphics[width=2.1in,angle=-90]{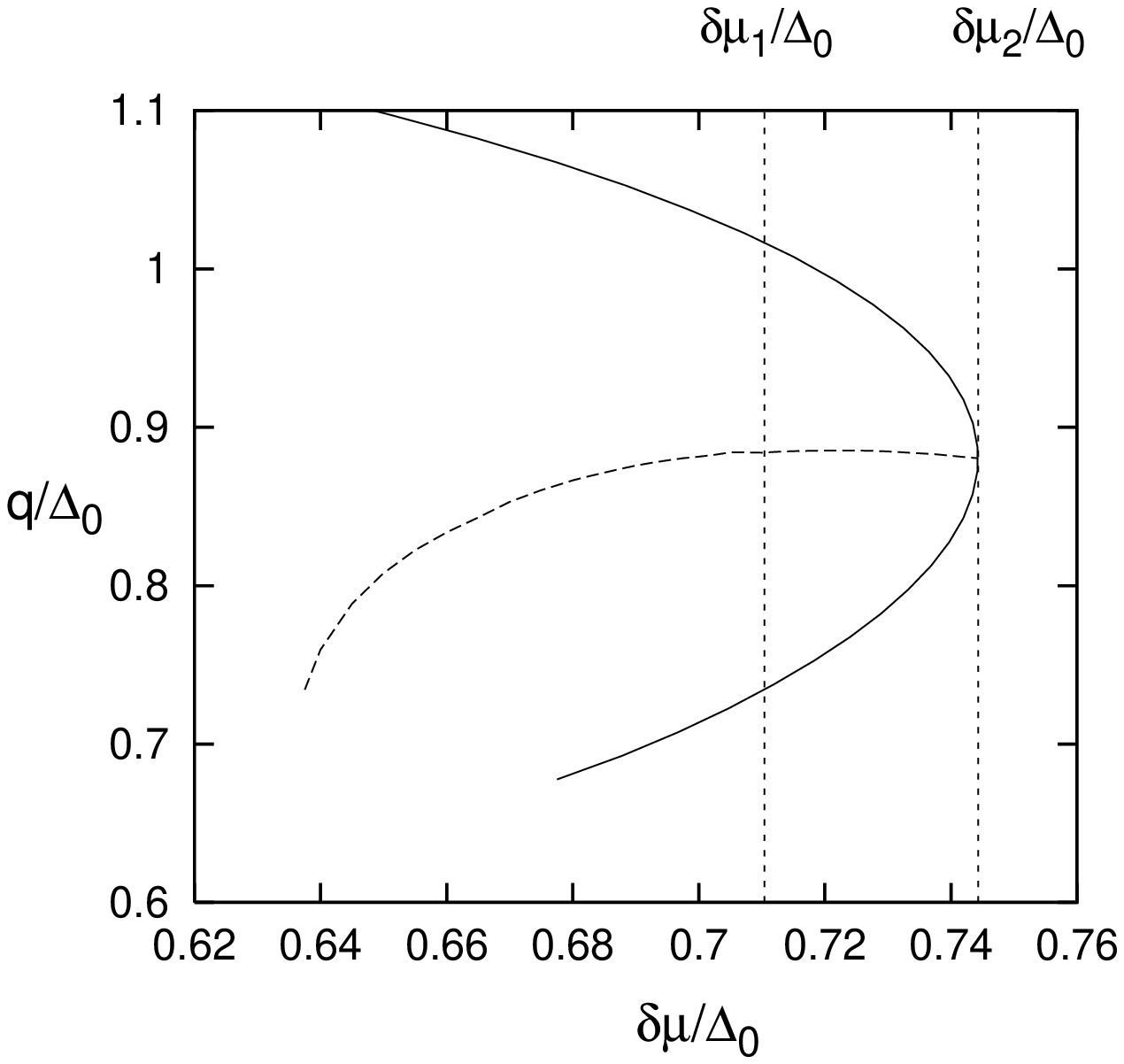}
\end{center}
\caption{
(a)~The zero-gap curves for the LOFF state. 
To the right of a solid curve, there is no solution to the LOFF
gap equation, to the left of the curve there is a solution,
and on the curve the gap parameter is zero.
The three curves are (from strongest
to weakest coupling): 
$\De_0=0.1, 0.04, 0.01~\GeV$.
The region $q<\dm$ is complicated to describe~\cite{FF},
and solutions found in this region never give the
lowest free energy state at a given $\dm$.
(b)~Here, we choose $\De_0=0.04~\GeV$
and focus on the region near $\dmmax$, the maximum value
of $\dm$ at which the LOFF state exists.   The dashed
curve shows the value of $|\vq|$ which minimizes the free energy of the LOFF
state at a given $\dm$.  $\dmmin$, discussed below, is also indicated.
}
\label{fig:zero_gap}
\end{figure}


We wish to determine $\dm_2$, the boundary separating
the LOFF phase and the normal phase.
The LOFF to unpaired phase transition is second order,
so it occurs where the solution $\De_A$ to the LOFF gap equation
\eqn{gap:LOFF} is zero.
Setting $\Delta_A=0$ in the gap equation (\ref{gap:BCSgap}) 
yields an analytical expression relating
$\dm$ and $q$, for any given $G$ and
$\La$.  In Figure~\ref{fig:zero_gap}a we show
the $\De_A=0$ curve
for three couplings corresponding to $\De_0=0.1~\GeV$ 
(strong coupling), $\De_0=0.04~\GeV$  
and $\De_0=0.01~\GeV$ (weak coupling).
We have only drawn the zero-gap curve in the region 
where $q\geq\dm$.
We expect this to be the region of interest for LOFF pairing
because when $q\geq\dm$ the two spheres of Figure~\ref{fig:regions}
do in fact intersect.  We have verified that, as described in some
detail in Ref.~\cite{FF}, there are regions of Figure~\ref{fig:zero_gap}a
with $q<\dm$ within which the LOFF gap equation (\ref{gap:LOFF})
has (one or even two) nonzero solutions, but these solutions
all correspond to phases whose free energy is either 
greater than that of the normal phase or greater than
that of the BCS phase or both.
Figure~\ref{fig:zero_gap} shows that 
for a given coupling strength, parameterized by $\De_0$, there is
a maximum $\dm$ for which the LOFF state exists: we call it
$\dmmax$. For $\dm>\dmmax$, the mismatch of chemical
potentials is too great for the LOFF phase to exist.

We see from Figure~\ref{fig:zero_gap}a that as the
coupling gets weaker, $\dmmax/\De_0$ gets gradually
larger. (Of course, $\dmmax$ itself gets smaller: 
the quantities plotted are $\dm/\De_0$ and $q/\De_0$.)
Note that in the $\De_0\rightarrow 0$ limit, the zero gap
curve  
is essentially that shown in the figure for $\De_0=0.01~\GeV$,
in agreement with the curve obtained at weak coupling
by Fulde and Ferrell~\cite{FF}.  The fact that this curve
ceases to move in the $\De_0\rightarrow 0$ limit means that
$\dmmax \rightarrow 0$ while $\dmmax/\De_0 \rightarrow {\rm const}$
in this limit.

For $\dm\rightarrow\dmmax$ from below, we see from 
Figure~\ref{fig:zero_gap} that
there is a solution to the LOFF gap equation only at a single
value of $q$.  For example, 
at $\De_0=0.04~\GeV$
we find $q=0.880\De_0=1.183\,\dmmax$ at $\dmmax=0.744\De_0$.
(In agreement with Refs.~\cite{LO,FF}, 
in the weak coupling limit we find
$q=0.906\De_0=1.20\,\dmmax$ at $\dmmax=0.754\De_0$.)
For any value of $\dm<\dmmax$,
solutions to the LOFF gap equation exist for a range of $|\vq|$.
We must now find the value of $|\vq|$ for which the free energy
of the LOFF state is minimized.   We obtain the
free energy of the LOFF state at a point in Figure~\ref{fig:zero_gap}
by first solving the gap equation (\ref{gap:LOFF}) numerically to obtain
$\Delta_A$, and then using (\ref{gap:DeltaDefn}) and (\ref{gap:tan2th})
to evaluate $\langle F_0 + H_I \rangle$ given in (\ref{gap:F0})
and (\ref{gap:HI}).  For
each value of $\dm<\dmmax$ we can now
determine which choice of $q$ yields the
lowest free energy.  The resulting ``best-$q$ curve'' curve  is shown in 
Figure~\ref{fig:zero_gap}b for $\De_0=0.04~\GeV$.\footnote{As a check
on our determination of the best $q$,  we have confirmed that
the total momentum of the LOFF state with the best $q$
is zero, as must be the case for the ground state of the
system at a given $\dm$ (by a theorem attributed to Bloch \cite{London}).   This is a powerful check, 
because it requires the 
net momentum of the unpaired quarks in the blocking regions
(which is in the negative $z$ direction; see Figure~\ref{fig:regions})
to be cancelled by the net momentum carried by the LOFF condensates.
When, in future work, our ansatz is extended to describe
a LOFF crystal rather than a single plane wave, this check
will no longer be powerful. Once we go from 
$\Gamma\sim\exp(2i\vq\cdot\vr)$ to $\Gamma\sim\cos(2\vq\cdot\vr)$
or to a more involved crystalline pattern, the total momentum
of the condensates and of the unpaired quarks will each be
zero.}

\begin{figure}[t]
\begin{center}
\includegraphics[width=2.38in,angle=-90]{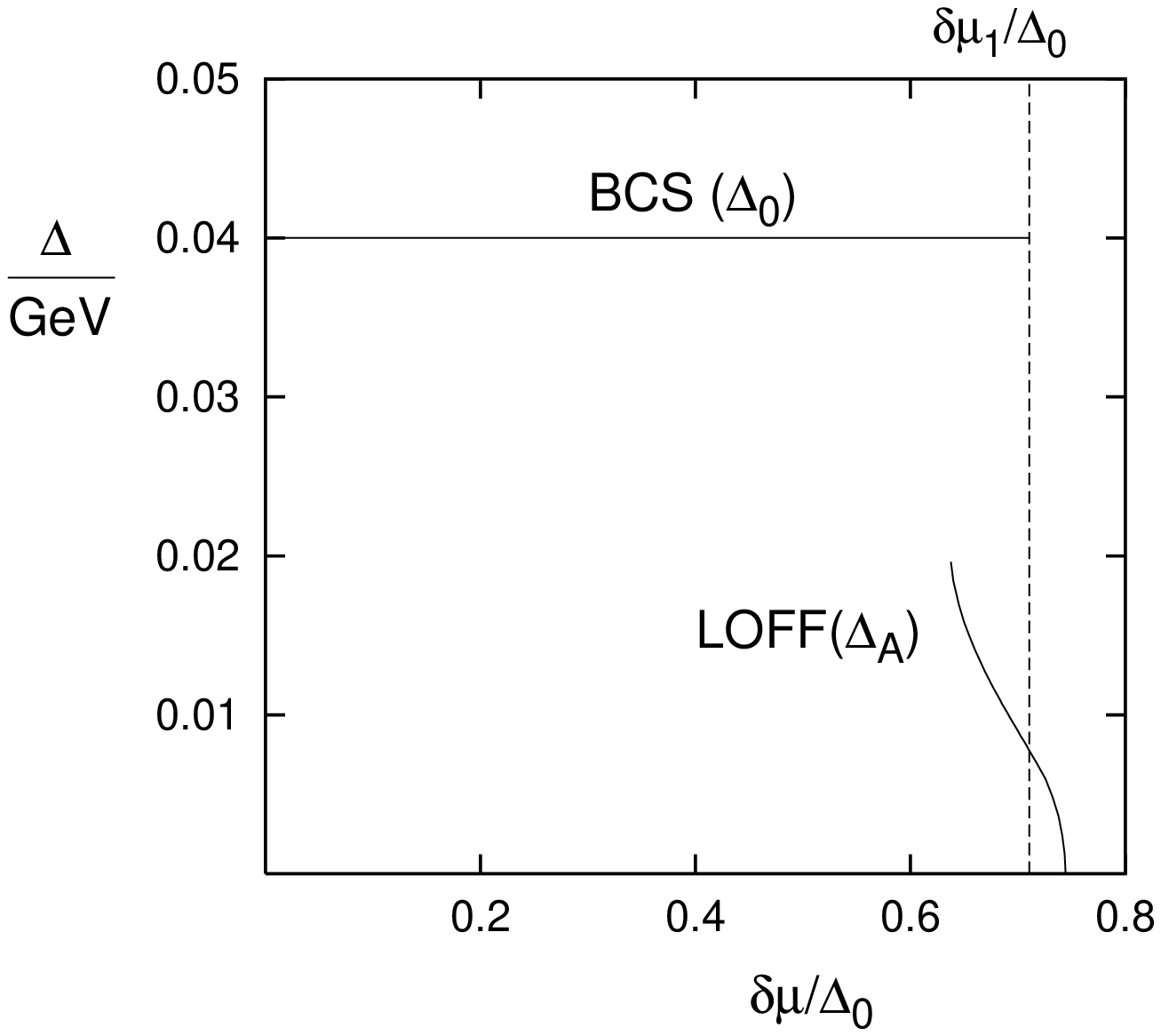}
\includegraphics[width=2.52in,angle=-90]{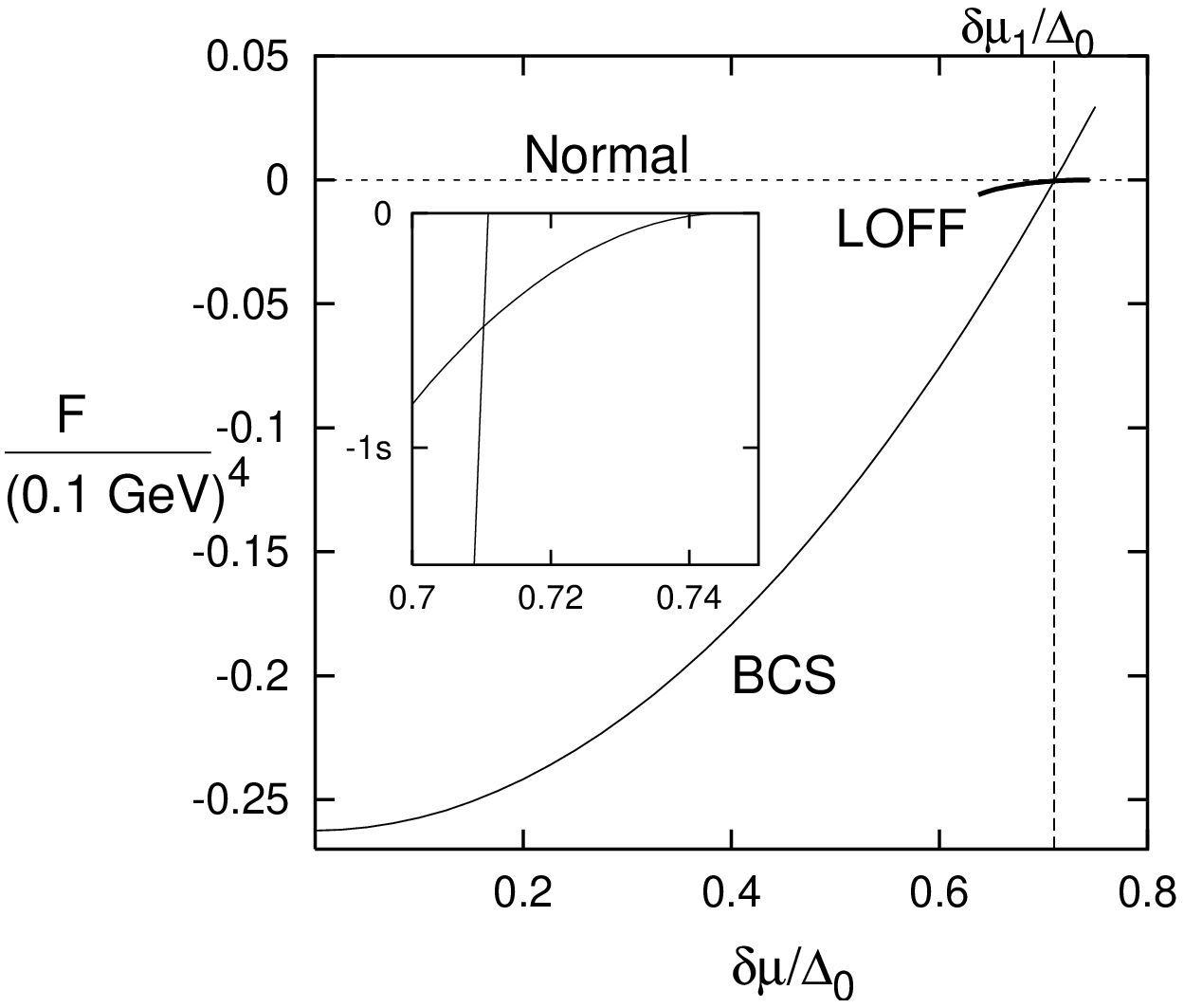}
\end{center}
\caption{LOFF and BCS gaps and free energies
as a function of $\dm$, with coupling chosen so 
that $\De_0=40{\rm ~MeV}$ and with $\mubar=0.4~\GeV,
\La=1~\GeV$. 
Free energies are measured relative to the normal state.
At each $\dm$ we have
varied $q$ to find the best LOFF state.
The vertical dashed line marks $\dm=\dm_1$, the value of $\dm$ above which
the LOFF state has lower free energy than BCS.
The expanded inset (wherein $s=10^{-7}~\GeV^4$) focuses on the region
$\dm_1<\dm<\dm_2$ where the LOFF state has the lowest free energy.
}
\label{fig:F_plot}
\end{figure}

Finally, for each point on the best-$q$ curve  
we ask whether the LOFF free energy at that $\dm$ and (best) $q$
is more or less than the
free energy of the BCS state at the same $\dm$. 
In this way, we find $\dmmin$ at which
a first order phase transition between the LOFF and BCS states
occurs.
In Figure~\ref{fig:F_plot} we show the competition between the
BCS and LOFF states as a function of the Fermi surface mismatch
$\dm$, for a fixed coupling corresponding to $\De_0=40$ MeV.  
The LOFF state exists for
$\dm<\dmmax=0.744\De_0$.  At each $\dm<\dmmax$, we plot the gap
parameter and free energy characterizing the LOFF state 
with the best $q$ for that $\dm$.
Although the BCS gap $\De_0$ is larger than the LOFF gap $\De_A$,
as $\dm$ increases we see from Eq.~(\ref{int:FBCS}) that
the BCS state pays a steadily increasing free-energetic price for maintaining
$p_F^u=p_F^d$, whereas the LOFF state pays no such price.
We now see that the LOFF state has lower free energy than the BCS state
for $\dm>\dmmin$, in this case $\dmmin=0.7104\De_0$.
At $\dm=\dmmin$, the gap parameter is $\De_A=0.0078~\GeV =0.195\De_0$. 
(Had we calculated $\dm_1$ by comparing the BCS free energy
with that of the unpaired state instead of with
that of the LOFF state, we would have obtained
$\dmmin=0.711\De_0$. 
As the inset to Figure~\ref{fig:F_plot} confirms,
the BCS free energy varies so rapidly that this makes an
almost imperceptible difference.  In later figures,
we therefore obtain $\dmmin$ via the simpler route of comparing
BCS vs. normal.)
At the coupling corresponding to $\De_0=40$ MeV, 
we have found that the LOFF state is favored over both the BCS state
and the normal state in a ``LOFF window''
$0.710 < \dm/\De_0 < 0.744$.

\begin{figure}
\begin{center}
\includegraphics[width=3in,angle=-90]{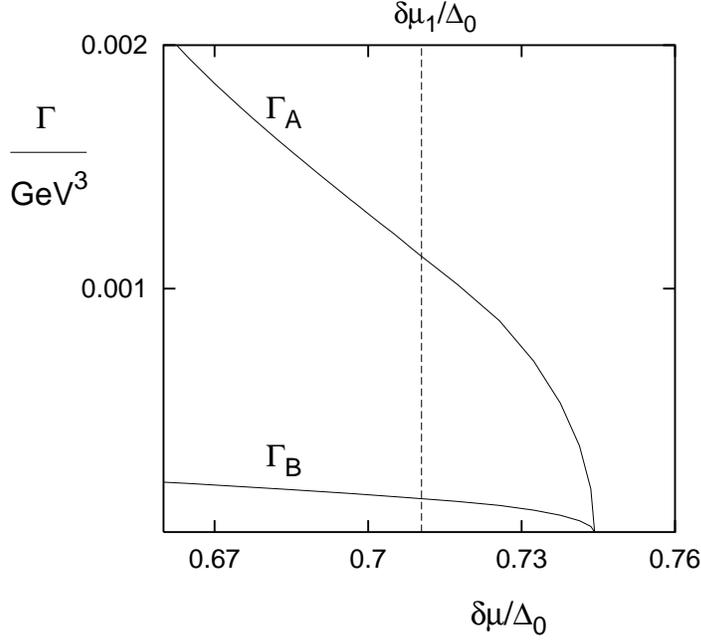}
\end{center} 
\caption{
The two LOFF condensates $\Gamma_A$ $(J=0)$
and $\Gamma_B$ $(J=1)$ for the same choice
of parameters as in Figure~\ref{fig:F_plot}. We focus on the region
$\dmmin<\dm<\dmmax$.   For reference, 
in the BCS phase $\Gamma_A=\De_0/G=0.00583~\GeV^3$ 
and $\Gamma_B=0$.
}
\label{fig:GammaBandA}
\end{figure}

With solutions to the gap equation in hand, we can obtain
the $J=0$ condensate $\Gamma_A=G\De_A$ and the $J=1$ condensate
$\Gamma_B$ given in Eq.~(\ref{gap:GammaB}). In Figure~\ref{fig:GammaBandA},
we show both condensates
within the LOFF window $\dmmin<\dm<\dmmax$.
We see first of all that $\Gamma_B\neq 0$, as advertised.
For the choice of parameters in Figs.~\ref{fig:F_plot} 
and \ref{fig:GammaBandA} we find 
$\Gamma_B/\Gamma_A$
essentially constant over the
whole LOFF window, varying from 0.121 at $\dmmin$ to 0.133 at $\dmmax$.
Increasing $\De_0$ tends
to increase $\Gamma_B/\Gamma_A$, as does decreasing $\La$.  
Second of all, we see that the phase transition at $\dm=\dmmax$,
between the LOFF and normal phases, is second order in the mean-field
approximation we employ throughout.

Near the second-order critical point $\dmmax$, we can describe the
phase transition with a Ginzburg-Landau effective potential.
The order parameter for the LOFF-to-normal phase transition is
\beql{lofforderparam}
\Phi(\vr) = -\frac{1}{2} \langle \epsilon_{ij} \epsilon_{\al\be3} \psi^{i\al}(\vr) C \gamma_5 \psi^{j\be}(\vr) \rangle 
\eeql
so that in the normal phase $\Phi(\vr) = 0$, while in the LOFF phase
$\Phi(\vr) = \Gamma_A e^{i 2 \vq \cdot \vr}$.  Expressing the order
parameter in terms of its Fourier modes $\tilde\Phi(\vk)$, we write
the LOFF free energy (relative to the normal state) as
\beql{ginzland}
F(\{\tilde\Phi(\vk)\}) = \sum_{\vk} \left( C_2(k^2) | \tilde\Phi(\vk) |^2 
+ C_4(k^2) | \tilde\Phi(\vk) |^4 + {\mathcal O}(|\tilde\Phi|^6) \right).
\eeql
For $\dm > \dmmax$, all of the $C_2(k^2)$ are positive and the normal
state is stable.  Just below the critical point, all of the modes
$\tilde\Phi(\vk)$ are stable except those on the sphere $|\vk| =
2q_2$, where $q_2$ is the value of $|\vq|$ at $\dmmax$ (so that $q_2
\simeq 1.2 \dmmax \simeq 0.9 \Delta_0$ at weak coupling).  In general,
therefore, many modes on this sphere can become nonzero, giving a
condensate with a complex crystal structure.  We consider the simplest
case of a plane wave condensate where only the one mode
$\tilde\Phi(\vk = 2\vq_2) = \Gamma_A$ is nonvanishing.  Dropping all
other modes, we have
\beql{ginzland2}
F(\Gamma_A) = a(\dm - \dmmax) (\Gamma_A)^2 + b (\Gamma_A)^4 
\eeql
where $a$ and $b$ are positive constants.  Finding the minimum-energy
solution for $\dm < \dmmax$, we obtain simple power-law relations
for the condensate and the free energy:
\beql{powerlaws}
\Gamma_A(\dm) = K_{\Gamma} (\dmmax - \dm)^{1/2}, \hspace{0.3in} F(\dm) = - K_F (\dmmax - \dm)^2.
\eeql
These expressions agree well with the numerical results shown in
Figs.~\ref{fig:F_plot} and \ref{fig:GammaBandA}.  The Ginzburg-Landau
method does not specify the proportionality factors $K_\Gamma$ and
$K_F$, but analytical expressions for these coefficients can be
obtained in the weak coupling limit by explicitly solving the gap
equation \cite{Takada1}, yielding
\beql{keqns}
\ba{rclcl}
G_A K_\Gamma &=& 2 \sqrt{\dmmax} \sqrt{(q_2/\dmmax)^2  - 1}  &\simeq& 1.15 \sqrt{\Delta_0} \\[0.5ex]
K_F &=& (4\bar\mu^2/\pi^2)((q_2/\dmmax)^2-1) &\simeq& 0.178 \bar\mu^2.
\ea
\eeql
Notice that because $(\dmmax-\dmmin)/\dmmax$ is quite small, the
power-law relations (\ref{powerlaws}) are a good model of the system
throughout the entire LOFF interval $\dmmin < \dm < \dmmax$ where the
LOFF phase is favored over the BCS phase.  The Ginzburg-Landau
expression (\ref{ginzland2}) gives the free energy of the LOFF phase
near $\dmmax$, but it cannot be used to determine the location
$\dmmin$ of the first-order phase transition where the LOFF window
terminates (locating the first-order point requires a comparison of
LOFF and BCS free energies).

\begin{figure}[t]
\begin{center}
\includegraphics[width=3in,angle=-90]{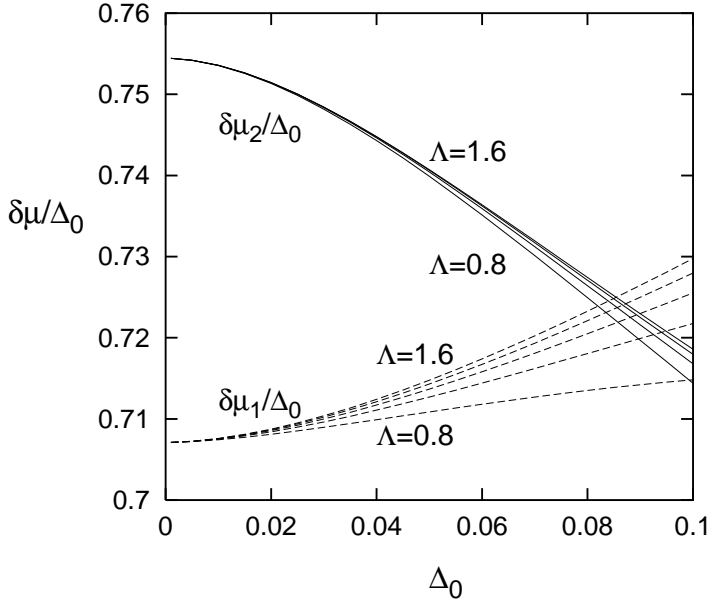}
\end{center}
\caption{The interval of $\dm$ within which the LOFF state occurs,
as a function of the coupling (parameterized as usual by the
BCS gap $\De_0$).
Below the solid line, there is a LOFF state. Below the dashed line,
the BCS state is favored. The different lines of each type correspond to
different cutoffs $\La=0.8~\GeV$ to $1.6~\GeV$. $\dmmin/\De_0$ 
and $\dmmax/\De_0$ show little
cutoff-dependence, and the cutoff-dependence disappears completely
as $\De_0,\dm\to 0$.
}
\label{fig:fish}
\end{figure}

It is interesting to explore how the width of the LOFF window
depends on the strength of the coupling, 
and to confirm that it is insensitive to the cutoff.
We do this in Figure~\ref{fig:fish}, where we plot $\dmmax/\De_0$
(solid lines) and $\dmmin/\De_0$ (dashed lines).
The LOFF state is favored
for $\dmmin/\De_0< \dm/\De_0 < \dmmax/\De_0$, i.e.~between the
solid and dashed curves in Figure~\ref{fig:fish}.
In the weak coupling limit, the LOFF window tends to 
$0.707 < \dm/\De_0 < 0.754$ and $\De_A$ at $\dmmin$ 
tends to $0.23\De_0$, as in Refs.~\cite{LO,FF}.  
Note that if one takes the weak-coupling limit $\De_0\rightarrow 0$
at fixed $\dm$, neither BCS nor LOFF pairing survives because
$\dm/\De_0\rightarrow\infty$.
However, for any arbitrarily
small but nonzero coupling, the LOFF phase
is favored within a range of $\dm$. 
Figure~\ref{fig:fish} thus demonstrates that in an analysis of the LOFF state 
in the weak-coupling limit, it is convenient to
keep $\dm/\De_0$ fixed while taking $\De_0\rightarrow 0$.
We see from Figure~\ref{fig:fish} 
that strong coupling helps the BCS state more than it
helps the LOFF state.
When the coupling gets strong enough, there is no longer
any window of Fermi surface mismatch $\dm$ in which the LOFF
state occurs: the BCS state is always preferred.

The different lines of each type in Figure~\ref{fig:fish} 
are for different cutoffs
and show that there is in fact little sensitivity to the cutoff.
The $\La$ dependence of $\dmmin/\De_0$ and $\dmmax/\De_0$ is mild
for all values of $\De_0$ which are of interest, 
and is weakest for $\De_0\rightarrow 0$. 
This is because in that limit pairing can only occur
very close to the unblocked ribbon of the ellipsoid of 
Fig. 2b, along which the integrand in the gap equation 
is singular and pairing is allowed. Thus
most of the pairing region ${\cal P}$,
and in particular the region near $\La$, become irrelevant in this limit.

The one physical quantity which we have explored which does
turn out to depend qualitatively on $\La$ is the ratio $\Gamma_B/\Gamma_A$.
Those quarks with 
momenta as large as $\La$ which pair have momenta which are 
almost antiparallel, and so contribute much less to $\Gamma_B$ than
to $\Gamma_A$.  For this reason, the ratio $\Gamma_B/\Gamma_A$
is sensitive to the number of Cooper pairs formed at very large $\vp$,
and hence to the choice of $\La$.  As discussed above, pairing
far from the favored ribbon in phase space becomes irrelevant
for $\De_0\rightarrow 0$, and indeed in this limit we find that the
$\La$ dependence of $\Gamma_B/\Gamma_A$ decreases.  However,
for $\De_0=40$ MeV we find that changing $\La$ from $1.2~\GeV$
to $0.8~\GeV$ increases $\Gamma_B/\Gamma_A$ by more than 50\%.

We chose to show results for $\De_0=40$ MeV in Figure~\ref{fig:F_plot} because
with this choice, the LOFF window occurs 
at values of $\dm$ comparable
to that in the illustrative example (\ref{int:illustrative}):
$\dm=\half(\mu_d-\mu_u) = 27$ MeV.
Of course, neither $\dm$  nor the value of $\De_0$ are accurately known
for the quark matter which may exist within a compact star.
Still, it seems possible that their ratio
could be appropriate for the quark matter to 
be in the LOFF phase.
If there is a range
of radii within a compact star in which quark matter
occurs with $\dmmin < \dm < \dmmax$, this quark matter
will be a crystalline color superconductor.

In Figure~\ref{fig:F_plot}, the LOFF gap parameter $\De_A$ is 
$7.8$ MeV at $\dm=\dmmin$.  It remains larger than 
typical neutron star temperatures $T_{\rm ns}\sim 1$ keV
until very close to $\dm=\dmmax$.  Similarly, the
LOFF free energy, which is
$4.8\times 10^{-8}~\GeV^4 =  4.8\times(10~\MeV)^4$
at $\dm=\dmmin$, 
is much larger than $T_{\rm ns}^4$ throughout the LOFF window
except very close to $\dm=\dmmax$.  
Furthermore, we shall see in 
Section~\ref{sec:glitch} that 
the free energy of the LOFF state is of the right order to 
lead to interesting 
glitch phenomena.

\section{More general Hamiltonian and ansatz}
\label{sec:ham}

In Section~\ref{sec:gap}, we introduced the four-fermion
interaction Hamiltonian $H_I$ of Eq.~(\ref{gap:ham}) with independent
couplings $G_E$ and $G_M$ for the interactions which model
the exchange of electric and magnetic gluons.
It proves convenient to use the linear combinations
\beql{ham:GAandGB}
\ba{rcl}
G_A &=& \frac{1}{4}(G_E + 3 G_M)\\[0.5ex]
G_B &=& \frac{1}{4}(G_E - G_M)\ ,
\ea
\eeql
of the coupling constants
in terms of which the expectation value of $H_I$ 
in the LOFF state (\ref{LOFF:ansatz}) becomes
\beql{ham:HI}
\langle H_I \rangle = - \half 
G_A V \left( |\Gamma^L_A|^2 + |\Gamma^R_A|^2 \right)
- \half G_B V \left( |\Gamma^L_B|^2 + |\Gamma^R_B|^2 \right)\ .
\eeql
Thus, a positive coupling $G_A$ describes an attractive interaction
which induces a $J=0$ condensate $\Gamma_A$.  As we have seen,
in the LOFF state this is necessarily accompanied by a $J=1$ condensate
$\Gamma_B$.  
In our analysis to this point, we have set $G_A=G>0$ and $G_B=0$.
We now discuss the general case, in which $G_B\neq 0$.

Before beginning, let us consider how to choose $G_B/G_A$ in order 
for our model Hamiltonian to be a reasonable toy model for
QCD at nonzero baryon density.  At zero density, of course,
Lorentz invariance requires $G_B=0$.  At high densities, on
the other hand, electric gluons are screened while 
static magnetic gluons are not.  (Magnetic gluons with nonzero
frequency are damped.)  We now know~\cite{Son} that at asymptotically
high densities it is in fact the exchange of magnetic gluons 
which dominates the pairing interaction.  This suggests
the choice $G_E=0$, corresponding to $G_B/G_A = -1/3$.   
At the accessible densities of interest to us, it is presumably
not appropriate to neglect $G_E$ completely.  Note also that
the four-fermion interaction induced by instantons in QCD
only yields interactions in flavor-antisymmetric channels.
It results in an attractive interaction in the $J=0$ channel and no
interaction in the 
$J=1$ channel.  Thus, although the instanton interaction cannot
be written in the form (\ref{gap:ham}), for our purposes it
can be thought of as adding a contribution to $G_A$, but none
to $G_B$.  
Hence our model is likely to best represent high density QCD
for a ratio of couplings lying somewhere in the range
\beql{ham:range}
-\frac{1}{3} < \frac{G_B}{G_A} < 0 \ .
\eeql 
We plot our results over a wider range of couplings below.

Once $G_B\neq 0$ and there is an interaction in the $J=1$ channel,
we expect, in addition to the $J=1$
condensate $\Gamma_B$, a $J=1$ gap parameter $\Delta_B$.
The quasiparticle dispersion relations
are then determined by $\Delta_A$ and $\Delta_B$, 
which are defined as
\beql{ham:Deltas}
\ba{rcl}
\Delta_A &=& G_A \Gamma_A\\
\Delta_B &=& G_B \Gamma_B \ .
\ea
\eeql
Following through the variational calculation as in
Section~\ref{sec:gap} leads to the coupled gap equations:
\beql{ham:coupledgapeqs}
\ba{rcl}
\Delta_A &=& \dsp\frac{2G_A}{V} \sum_{\vp \in {\cal P}} 
\frac{ 2 S_A 
(\Delta_A S_A + \Delta_B S_B) }
{ \sqrt{ ( |\vq+\vp| + 
|\vq-\vp| - 2\bar\mu)^2 + 4 (\Delta_A S_A 
+ \Delta_B S_B)^2 }} \\[2ex]
\Delta_B &=& \dsp\frac{2G_B}{V} \sum_{\vp \in {\cal P}} 
\frac{ 2 S_B
(\Delta_A S_A + \Delta_B S_B) }
{ \sqrt{ ( |\vq+\vp| + 
|\vq-\vp| - 2\bar\mu)^2 + 4 (\Delta_A S_A 
+ \Delta_B S_B)^2 }} \\[3ex]
S_A &=& \sin(\half\beta_A(\vp)) \\[1ex]
S_B &=& \sin(\half\beta_B(\vp))
\ea
\eeql
with $\beta_A(\vp) = \alpha_u(\vp) + 
\alpha_d(\vp)$, $\beta_B(\vp) = \alpha_u(\vp) - \alpha_d(\vp)$
defined in terms of the angles in Figure~\ref{fig:angles}.
The pairing region ${\cal P}$ is still defined 
by \eqn{gap:pregion} but with new 
quasiparticle dispersion relations obtained from
Eqs.~(\ref{gap:quasi}) with $\De_A^2S_A^2$ replaced by
$(\De_A S_A + \De_B S_B)^2$.

For $G_B=0$, the coupled equations \eqn{ham:coupledgapeqs}
reduce to Eqs.~(\ref{gap:LOFF}) and
(\ref{gap:GammaB}).  Note that if, instead, $G_B>0$ and $G_A=0$, 
we find an attractive interaction
in the $J=1$ channel in Eq.~(\ref{ham:GAandGB}) and
no interaction in the $J=0$ channel.  Analysis
of  Eqs.~(\ref{ham:coupledgapeqs}) in this case yields a nonzero
value of $\Delta_B$, while $\Delta_A=0$ even though $\Gamma_A\neq 0$.
The geometry of the LOFF pairs  
requires $\Gamma_A\neq 0$ when $\Gamma_B \neq 0$.

Rather than describing how every Figure in Section~\ref{sec:results}
changes when $G_B\neq 0$, we choose to focus on the question
of how the interval of $\dm$ within which the LOFF state occurs
(the LOFF window)
changes as a function of $G_B/G_A$.  To further simplify the
presentation, we specialize to the weak-coupling limit
in which $\Delta_0\rightarrow 0$. This means that, as in 
Figure~\ref{fig:fish}, the LOFF window is independent of the
cutoff $\Lambda$.  

\begin{figure}[t]
\begin{center}
\includegraphics[width=3in,angle=-90]{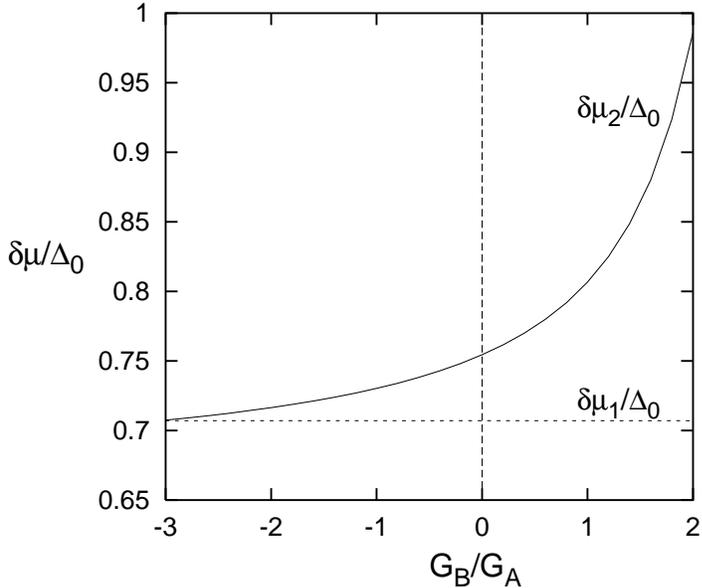}
\end{center} 
\caption{
The interval of $\dm$ in which the LOFF state 
is favored at weak coupling, as a function of the
ratio of couplings $G_B/G_A$. 
Below the solid line, there is  a LOFF state.
Below the dashed line, the ordinary BCS state is favored.
$G_B=0$ corresponds to the Lorentz-invariant interaction with $G_E=G_M$.
QCD at high density is likely best described by a coupling
in the range $-\third < G_B/G_A < 0$.
}
\label{fig:dm_vs_ba}
\end{figure}

We show the dependence of the LOFF window on $G_B/G_A$ in
Figure~\ref{fig:dm_vs_ba}. The lower boundary $\dm=\dm_1$ is, as in
Section~\ref{sec:results}, the same (up to a very small correction) as
the $\dm$ at which the BCS and normal states have equal free energies.
We find the upper boundary $\dm=\dm_2$
by first dividing Eqs.~(\ref{ham:coupledgapeqs}) by $\Delta_A$
and then looking for a value of $\dm$ at which $\Delta_A\rightarrow 0$
and $\Delta_B\rightarrow 0$ but $\Delta_A/\Delta_B$ remains nonzero.
As before, this defines a zero-gap curve, and $\dmmax$ is the
maximum value of $\dm$ reached by this curve.

%
%

We find that the lower boundary $\dm_1$ is completely unaffected by
the value of $G_B$, since the BCS state is purely $J=0$. So in the
weak-coupling limit we obtain the result of Section~\ref{sec:results},
$\dm_1/\De_0=0.707$, independent of $G_B/G_A$.  In contrast, $\dm_2$,
the upper boundary of the LOFF window, increases with increasing
$G_B$. This is understandable: the LOFF state always produces a $J=1$
condensate, so we expect it to be fortified by $G_B>0$ and penalized
by $G_B<0$.  
There is no analogue of this behavior in an electron
superconductor~\cite{LO,FF}, where there can
be no $J=1$ condensate.  Our $J=1$ condensate
affects the gap equation and free energy 
only if $G_B\neq 0$;  for this reason, our weak coupling results
are in agreement
with those of LOFF~\cite{LO,FF} only if $G_B=0$, as in 
Section~\ref{sec:results}.
The effect of a coupling $G_B$
in the physically interesting range (\ref{ham:range}) is to reduce the
LOFF window, but only slightly.



In both this Section and the previous one, we have calculated $\dm_2$
by examining the competition between LOFF pairing and no pairing.
Should we instead have considered the competition between LOFF pairing
and the formation of $\langle uu\rangle$ and $\langle dd\rangle$
condensates, each at their respective Fermi surface, each with
$\vq=0$?  Assuming as usual that the color antisymmetric channel is
the most attractive one, flavor symmetric pairing requires
spin-symmetric pairing, i.e.~$J=1$~\cite{TS1flav}.  Within the model
and ansatz that we have considered, the question is easily answered.
If we choose $G_A>0$ and $G_B=0$, as in Sections~\ref{sec:LOFF} to
\ref{sec:results}, there is no interaction in the spin-symmetric,
flavor-symmetric, color-antisymmetric channel.  If we strengthen
magnetic gluon exchange relative to electric gluon exchange by
choosing $G_B<0$, the interaction in this channel is repulsive.  We
have confirmed this by evaluating the expectation value of $H_I$ in a
state with spatially uniform $J=1$ pairing and $\langle uC\sigma^{0i}u
\rangle$ condensate (obtained by using two $u$ creation operators in
the ansatz (\ref{LOFF:ansatz}), setting $\dm=0$ as appropriate for
$\langle uu \rangle$ pairing, setting $\vq=0$ and removing the
blocking regions).  We find that $G_A$ gives no interaction in this
channel and $G_B<0$ is repulsive.  Thus, for the same reason that the
LOFF window shrinks for $G_B<0$, there can be no $\langle uu\rangle$
or $\langle dd\rangle$ pairing.  However, the scenario is apparently
different at asymptotically high density: it has been shown by
Sch\"afer \cite{TS1flav} that long-range single-gluon exchange does in
fact induce pairing in this $J=1$ channel.  (The long-range interaction
emphasizes near-collinear scattering which is attractive for both
electric and magnetic gluons.)  For either a pointlike interaction
with $G_B>0$ or a long-range interaction dominated by near-collinear
scattering, we therefore expect competition between LOFF pairing and
$\langle uu\rangle$ and $\langle dd\rangle$ pairing, since the latter
would then be favored for $\dm>\dm_2$.

Our ansatz only contains $LL$ and $RR$ pairing.  
We leave a complete analysis of 
the generalization to $LR$ pairing to future work.
We have, however, constructed the ansatz for 
spatially uniform $LR$ 
pairing with $\dm=0$ and $\vq=0$.
We find that the interaction in this $J=1$ channel is
attractive if $G_E + G_M > 0$ and is independent 
of the linear combination of couplings $G_E -3 G_M$.  The 
$J=1$ channel with LR pairing yields a 
$\langle u C \gamma^i u \rangle$ condensate instead of the
$J=1$ condensate 
$\langle u C \sigma^{0i} u \rangle$ obtained for the case of $LL$ 
and $RR$ pairing.  In agreement with Ref.~\cite{TS1flav}
we find that magnetic
gluon exchange,  with $G_E=0$ and $G_M >0$, 
is attractive in the $\langle u C \gamma^i u \rangle$
channel.
Note that in the nonrelativistic limit
$\langle u C \gamma^i u \rangle$ and $\langle u C \sigma^{0i} u \rangle$
are equivalent $J=1$ condensates.  In the relativistic
setting relevant in quark matter, we find that pointlike
interactions in these two channels have opposite sign.

We have set up the
gap equation describing a spatially uniform $\langle u C \gamma^i u \rangle$
condensate
and solved it for $G_E=G_M=G$,
$\De_0=40~\MeV$, $\mu_u=0.4~\GeV$, and $\La=1~\GeV$.
We find a gap of 8 keV and a free energy
which is about five orders of magnitude
smaller than that of the LOFF phase.  (If we
choose $G_E=0$ and $G_M>0$, the interaction is
still attractive but the gap is even smaller.)  
Therefore, even though for $\dm>\dmmax$
we expect LR pairing and consequent
$\langle u C \gamma^i u \rangle$
and $\langle d C \gamma^i d \rangle$ condensates, 
the resulting condensation
energy is so small that it is a good approximation
to neglect these condensates in the evaluation of $\dmmax$, as we
have done.
We leave for future work
a complete analysis of
the competition between the LOFF phase (with an 
ansatz extended to allow $LR$-LOFF pairing) and
the spatially uniform $\langle u C \gamma^i u \rangle$ 
condensate.

\section{Conclusions, future work and astrophysical implications}
\label{sec:conclusions}

\subsection{Conclusions}

We have studied the formation of a
rotational-symmetry-breaking LOFF state involving pairing between
two flavors of quark whose chemical potentials differ by $2\dm$.
This state is characterized by a gap parameter and a diquark condensate,
but not by an energy gap in the dispersion relation.  
In the LOFF state, each Cooper pair carries momentum $2\vq$ 
with $|\vq| \approx 1.2\dm$.
The condensate and gap parameter vary in space with wavelength
$\pi/|\vq|$.

We focused primarily on
an NJL-type four-fermion interaction
with the quantum numbers of single gluon exchange.
In the limit of weak coupling (BCS gap $\De_0\ll \mu$)
the LOFF state is favored for
values of $\dm$ which satisfy
$\dm_1 < \dm < \dm_2$, where $\dm_1/\De_0=0.707$ and 
$\dm_2/\De_0=0.754$.
The LOFF gap parameter decreases from $0.23 \De_0$
at $\dm=\dm_1$ to zero at $\dm=\dm_2$.  
These are the same results found by LOFF in their original analysis.
Except for very close to $\dm_2$, the critical
temperature above which the LOFF state melts will be much
higher than typical neutron star temperatures.
At stronger coupling the LOFF gap parameter decreases relative
to $\De_0$ and 
the window of $\dm/\De_0$ within which the LOFF state
is favored shrinks. The window grows
if the interaction is changed to weight electric
gluon exchange more heavily than magnetic gluon exchange.

Because it violates rotational
invariance by involving Cooper pairs whose
momenta are not antiparallel, the quark matter LOFF state  
necessarily features nonzero condensates in 
both the $J=0$ and $J=1$ channels.
Both condensates are present even if there is no
interaction in the $J=1$ channel.  In this case, 
however, the $J=1$ condensate does not affect the
quasiparticle dispersion relations; that is, the $J=1$
gap parameter vanishes.  If there is an attraction in
the $J=1$ channel (as, for example, if the strength
of the electric gluon interaction is increased) the 
size of the LOFF window increases.

The quark matter which may be present within a compact star will be in
the crystalline color superconductor (LOFF) state 
if $\dm/\De_0$ is in the requisite range.  For $\dm$ as in
the illustrative example (\ref{int:illustrative}), this
occurs if the gap $\De_0$ which characterizes the uniform
color superconductor present at smaller values of $\dm$ is 
about 40 MeV. This is in the middle of the range of present
estimates.  Both $\dm$ and $\De_0$ vary as a function
of density and hence as a function of radius in a compact star.
Although it is too early to make quantitative predictions,
the numbers are such that crystalline color superconducting
quark matter may very well occur in a range of radii within a compact 
star. It is therefore worthwhile to consider the consequences.

\subsection{Future work}
\label{sec:future}

The prospect of spontaneous violation of translational and rotational
symmetry in dense quark matter is very exciting.  In the final pages
of this paper we will begin to explore one particularly
interesting consequence: glitch behavior in quark matter within compact
stars.  First, however, we list a number of direct extensions
of our work, several of which are prerequisites to a quantitative
exploration of the astrophysical consequences of crystalline
color superconducting quark matter in compact stars.

\begin{enumerate}
\setlength{\itemsep}{-0.5\parsep}
\item We have restricted ourselves to two flavors of quark,
and varied $\dm$ freely.
It is crucial to look at more realistic examples, imposing charge
neutrality and weak equilibrium, and including the strange
quark.  We expect a LOFF phase wherever $\langle us \rangle$,
$\langle ds \rangle$ or $\langle ud \rangle$ pairs approach their
unpairing transitions, but this must be verified quantitatively.
Further generalizations would
include bare quark masses and spontaneous generation
of constituent quark masses by chiral condensation.
\item It would be valuable to complement our NJL-model study with a
controlled calculation using one gluon exchange in the asymptotically
high density limit.  There are two reasons why this is
worthwhile. First, it will allow a controlled analysis without model
assumptions, albeit one of quantitative value only at extremely high
densities.  In particular, this would allow a better estimation of the
relative magnitude of the $J=1$ and $J=0$ condensates, which was the
one feature which we found to depend strongly on the choice of cutoff
in our model.  Second, quark-quark scattering by the exchange of a
gluon at weak coupling is dominated by small-angle scattering, whereas
in an NJL model of the type we have used this is not the case.  This
can actually affect the sign of the interaction in the $J=1$ channel
and perhaps thereby increase the range of the LOFF window, as we
pointed out in the previous section.  Moreover, it is known that the
LOFF window is much wider in one dimension than in three~\cite{1D},
and since the three-dimensional physics at asymptotically high
densities can be treated as a sum of one-dimensional
theories~\cite{Hong,BBS}, we have another reason to suspect that the
LOFF window may be wider at asymptotically high densities than our
present analysis would suggest.
\item As we have discussed at length in Section~\ref{sec:ham},
it would be of interest to extend our treatment to include
pairing between quarks of the same flavor and pairing between
quarks of opposite chirality.

\item Perhaps the most crucial unresolved issue
is the question of what crystal structure the LOFF phase
chooses.  Larkin and Ovchinnikov concluded that the condensate
varies in space like \mbox{$\cos(2\vq\cdot\vr)$}, 
forming a one-dimensional
standing wave with nodal planes spaced every $\pi/(2|\vq|)$.
The competition between this planar structure and one with, say,
a cubic or body-centered-cubic crystal structure is subtle.
In two dimensions, the answer depends sensitively on 
the temperature~\cite{Shimahara}; in three dimensions, 
it is apparently still unresolved even in the 
original LOFF context~\cite{Buzdin}.
In the QCD context, with the added complication of a $J=1$
condensate, it will be quite interesting to determine
what pattern is favored.

\item Finally, it would be very interesting to investigate the
astrophysical consequences of the LOFF phase. Since it occurs
in a range of $\dm$, one would expect that quark matter
stars could contain a layer of crystalline
LOFF condensate. In the next subsection, we take some
preliminary steps in this investigation.

\end{enumerate}

\subsection{Looking ahead to astrophysical consequences}
\label{sec:glitch}

Many pulsars have been observed to glitch.  Glitches are sudden
jumps in rotation frequency $\Omega$ which may
be as large as $\Delta\Omega/\Omega\sim 10^{-6}$, but may also
be several orders of magnitude smaller.  The frequency of observed
glitches is statistically consistent with the hypothesis that 
all radio pulsars experience glitches~\cite{AlparHo}.
Glitches are thought to originate from interactions
between the rigid crust, somewhat more than a kilometer thick in a typical
neutron star, and rotational vortices in the 
neutron superfluid.  
The inner kilometer of the crust
consists of a rigid lattice of nuclei immersed in 
a neutron superfluid~\cite{NegeleVautherin}.
Because the pulsar is spinning, the neutron superfluid 
(both within the inner crust and deeper inside the star) 
is threaded with
a regular array of rotational vortices.  As the pulsar's spin
gradually slows due to emission of electromagnetic radiation,
these vortices must gradually move outwards since the rotation frequency
of a superfluid is proportional to the density of vortices. 
Deep within the star, the vortices are free to move outwards.
In the crust, however, the vortices are pinned by their interaction
with the nuclear lattice.  What happens next varies from model to
model.  Perhaps the vortices exert sufficient force on the crust
to tear it apart, resulting in a sudden breaking and rearrangement of
the crust and a change in the moment of inertia~\cite{RudermanGlitch}.
Perhaps a large cluster of vortices within the inner crust
builds up enough outward pressure to overcome the pinning force,
suddenly becomes unpinned, and moves macroscopically 
outward~\cite{AndersonItoh,AndersonEtAl,AAPS1,AAPS2,AAPS3,PinesAlpar,Recent}.
This sudden decrease in the angular momentum
of the superfluid within the crust results in a sudden increase
in angular momentum of the rigid crust itself, and hence a glitch.
Perhaps, due to interactions between neutron
vortices and proton flux tubes, the neutron vortices pile up
just inside the inner crust before suddenly coming 
unpinned~\cite{SedrakianCordes}.  Although the models differ
in important respects, all agree that the fundamental requirements
are the presence of rotational vortices in a 
superfluid 
and the presence
of a rigid structure which impedes the motion of vortices and
which encompasses enough of the volume of the pulsar to contribute
significantly to the total moment of inertia.\footnote{The
first model of glitches which was proposed~\cite{Starquake}
relies on the cracking and settling
of the neutron star crust (``starquakes'') as the neutron star spins down.
This model does not require the presence of rotational vortices.
However, this model fails to explain the magnitude and
frequency of glitches in the Vela pulsar~\cite{PinesAlpar,Recent}.}

Although it is 
premature to draw quantitative conclusions,
it is interesting to speculate that some glitches may originate 
deep within a pulsar which features
a quark matter core, in a region of that core 
in which the color superconducting quark matter is in
a LOFF crystalline color superconductor phase.
The first prerequisite for
a quantitative answer to whether this may occur is to repeat
our analysis in the more general context of three-flavor quark
matter with a nonzero strange quark mass
$M_s$, to estimate over what range
of densities LOFF phases may arise, as either 
$\langle ud \rangle$, $\langle us \rangle$ or $\langle ds \rangle$
condensates approach their unpairing transitions.  Comparison
to existing models which describe how $p_F^u$, $p_F^d$ and $p_F^s$
vary within a quark matter core in a neutron star~\cite{Glendenning} 
would then
permit an estimate of how much the LOFF region contributes to
the moment of inertia of the pulsar.  Furthermore, a three 
flavor analysis is required to determine whether the LOFF
phase is a superfluid.   If the only pairing is between $u$
and $d$ quarks, this 2SC phase is not a superfluid~\cite{ARW2,ABR2+1},
whereas if all three
quarks pair in some way, a superfluid {\it is} 
obtained~\cite{ARW3,ABR2+1}.\footnote{As 
an aside, note that the crystalline 
chiral condensate~\cite{RappCrystal} (due to particle-hole pairing 
which may form at sufficiently strong coupling or at very large $N_c$)
is not a superfluid.}

Henceforth, we suppose  that the LOFF phase is a superfluid, 
which means that if it occurs within a pulsar it will be threaded
by an array of rotational vortices.
It is reasonable to expect that these vortices will
be pinned in a LOFF crystal, in which the
diquark condensate varies periodically in space.  
Indeed, one of the suggestions for how to look for a LOFF phase in
terrestrial electron superconductors relies on the fact that
the pinning of magnetic flux tubes (which, like the rotational vortices
of interest to us, have normal cores)
is expected to be much stronger
in a LOFF phase than in a uniform BCS superconductor~\cite{Modler}.

A real calculation of the pinning force experienced by a vortex in a
crystalline color superconductor must await the determination of the
crystal structure of the LOFF phase. We can, however, attempt an order
of magnitude estimate along the same lines as that done by Anderson
and Itoh~\cite{AndersonItoh} for neutron vortices in the inner crust
of a neutron star. In that context, this estimate has since been made
quantitative~\cite{Alpar77,AAPS3,Recent}.  With parameters chosen  
as in Figure~\ref{fig:F_plot}, we find that at $\dm=\dmmin$
the LOFF phase is favored over the normal state
by a free energy $F_{\rm LOFF}\sim
5 \times (10 {\rm ~MeV})^4$ and the spacing between nodes in the LOFF
crystal is $b=\pi/(2|\vq|)\sim 9$ fm.  
The thickness of a rotational vortex is
given by the correlation length $\xi\sim 1/\Delta_A \sim 25$ fm.  
All these numbers
are quite uncertain, but we will use them for the present.
In the context of crustal
neutron superfluid vortices, there are three
distinct length scales: the vortex thickness $\xi$, the lattice
spacing between nuclei $b$, and $R$, the radius of the individual
nuclei.  (The condensate vanishes within regions of size $R$ separated
by spacing $b$.)  In the LOFF phase, the latter two length scales are
comparable: since the
condensate varies like $\cos(\pi r/b)$ it is as if $R\sim b$.  The
fact that these length scales are similar in the LOFF phase will
complicate a quantitative calculation of the pinning energy; it makes
our order of magnitude estimation easier, however.  The pinning energy
is the difference between the energy of a section of vortex of length 
$b$ which is centered on a node of the LOFF crystal vs. one which
is centered on a maximum of the LOFF crystal. It 
is of order
\begin{equation}
E_p \sim F_{\rm LOFF}\, b^3 \sim 4 {\rm \ MeV}\ .
\end{equation}
The resulting pinning force per unit length of vortex is of order
\begin{equation}
f_p \sim \frac{E_p}{b^2} \sim  \frac{4 {\rm \ MeV}}{80 {\rm \ fm}^2}\ .
\label{fpLOFF}
\end{equation}
A complete calculation will be challenging because
$b<\xi$, and is likely to yield an $f_p$
which is somewhat less than that we have obtained by dimensional 
analysis~\cite{AAPS3,Recent}.
Note that our estimate of $f_p$ is
quite uncertain both because it is
only based on dimensional analysis and because the values
of $\Delta_A$, $b$ and $F_{\rm LOFF}$ are 
uncertain.  (We know the values of
all the ratios $\Delta_A/\Delta_0$, $\dm/\De_0$, $q/\De_0$ 
and consequently $b\De_0$ quite accurately in the LOFF phase.  It is 
of course the value of the BCS gap $\De_0$ which is uncertain.) 
It is therefore premature to compare our crude result 
to the results of serious calculations 
of the pinning of crustal neutron vortices as in 
Refs.~\cite{Alpar77,AAPS3,Recent}.  It is nevertheless
remarkable that they prove to be similar: the pinning
energy of neutron vortices in the inner crust 
is~\cite{AAPS3}
\begin{equation}
E_p \approx 1-3  {\rm \ MeV}
\end{equation}
and the pinning force per unit length is~\cite{AAPS3,PinesAlpar}
\begin{equation}
f_p\sim \frac{E_p}{b\xi}\approx
\frac{1-3 {\rm ~MeV}}{(25-50 {\rm ~fm})(4-20{\rm ~fm})}\ ,
\end{equation}
where the form of this expression is appropriate because $\xi<b$.
Perhaps, therefore, glitches occurring in a region of crystalline
color superconducting quark matter may yield similar phenomenology
to those occurring in the inner crust.

The reader
may be concerned that a glitch deep within the quark
matter core of a neutron star may not be observable:  the
vortices within the crystalline
color superconductor region suddenly unpin and leap
outward;  this loss of angular momentum is compensated
by a gain in angular momentum of the layer outside the
LOFF region; how quickly, then, does this increase
in angular momentum manifest itself at the {\em surface} of
the star as a glitch? If the LOFF layer is the outer
layer of the quark matter core---not unreasonable
since the chemical potential differences will be larger 
here than deeper inside the quark matter---there is
no problem.  The LOFF glitch speeds up the nucleon superfluid
outside the quark matter core, and the rotation
of this superfluid is coupled to
the rotation of the outer
crust on very short time scales~\cite{AlparLangerSauls}.
This rapid coupling, due to electron scattering off vortices
and the fact that the electron fluid penetrates throughout the 
star, is usually invoked to explain that the core
nucleon superfluid speeds up quickly after a crustal glitch:
the only long relaxation time is that of the vortices within
the inner crust~\cite{AlparLangerSauls}.
Here, we invoke it to explain that the outer crust speeds
up rapidly after a LOFF glitch has accelerated the quark matter
at the base of the nucleon superfluid. 
After a glitch in the LOFF region, the only
long relaxation times are those of the vortices in the LOFF
region and in the inner crust.

A quantitative theory of glitches originating within
quark matter in a LOFF phase must await the further
microscopic calculations sketched in Section~\ref{sec:future}. In
particular, an understanding of points 1 and 4 of 
Section~\ref{sec:future}
is a mandatory prerequisite.
However, our 
rough estimate of the pinning force on rotational vortices
in a LOFF region suggests that this force may be
comparable in magnitude to that on vortices in the inner
crust of a conventional neutron star, which yields glitches
in accord with those observed in pulsars.  This is surely
strong motivation for further investigation.

Perhaps the most interesting consequence of these speculations
arises in the context of compact stars made entirely of 
strange quark matter.  The work of Witten~\cite{Witten}
and Farhi and Jaffe~\cite{FarhiJaffe} raised the possibility
that strange quark matter may be energetically stable relative
to nuclear matter even at zero pressure.  If this is the
case it raises the question whether observed compact stars---pulsars,
for example---are strange quark stars~\cite{HZS,AFO} rather than
neutron stars.  
A conventional neutron star may feature
a core made of strange quark matter, as we have been discussing above. 
Strange quark stars, on the other hand, are made (almost)
entirely of quark
matter with either no hadronic matter content at all or
with a thin crust, of order one hundred meters thick, which contains
no neutron superfluid~\cite{AFO,GlendenningWeber}.
The nuclei in this thin crust
are supported above the quark matter by electrostatic forces;
these forces cannot support a neutron fluid.  Because
of the absence of superfluid neutrons, and because of the thinness of
the crust, no successful models of glitches in the crust
of a strange quark star have been proposed.  
Since pulsars are observed to glitch, the apparent lack of a 
glitch mechanism for strange
quark stars  has been the 
strongest argument that pulsars cannot be strange quark 
stars~\cite{Alpar,OldMadsen,Caldwell}.
This conclusion must now be revisited.  

Madsen's conclusion~\cite{Madsen} that a strange
quark star is prone to r-mode instability due to
the absence of damping must
also be revisited, since the relevant fluid oscillations
may be damped within or at the boundary of a region
of crystalline color superconductor.

The quark 
matter in a strange quark star, should
one exist, would be a color superconductor.
Depending on the mass of the star, the 
quark number densities increase by a factor of about two to ten
in going from the surface to the center~\cite{AFO}. This means
that the chemical potential differences among the three
quarks will vary also, and there could be a range of radii
within which the quark matter is in a crystalline
color superconductor phase.  This raises the 
possibility of glitches in strange quark stars.
Because the
variation in density with radius is gradual, if a shell
of LOFF quark matter exists it need not be particularly thin.
And, we have seen, the pinning forces may be comparable
in magnitude to those in the inner crust of a conventional
neutron star.
It has recently been suggested (for reasons unrelated to our considerations)
that certain accreting compact stars
may be strange quark stars~\cite{Bombaci}, although the
evidence is far from unambiguous~\cite{ChakrabartyPsaltis}.
In contrast, 
it has been thought that, because they glitch,  
conventional radio pulsars cannot be strange
quark stars.  Our work questions this assertion
by raising the possibility that glitches
may originate within a layer of quark matter 
which is in a crystalline color superconducting state.

\vspace{3ex}
{\samepage 
\begin{center} Acknowledgements \end{center}
\nopagebreak
We are grateful to C. Nayak for pointing out 
Refs.~\cite{LO} and \cite{FF} to us at the Aspen Center for Physics
more than one year ago.
We are grateful to him and to P.~Bedaque, J.~Berges, I.~Bombaci,
D.~Blaschke, D.~Chakrabarty, R.~Jaffe,
J.~Madsen, D.~Psaltis, S.-J.~Rey, M.~Ruderman, T.~Sch\"afer, A.~Sedrakian,
E.~Shuster, D.~Son, M.~Stephanov,
I.~Wasserman, F.~Weber and F.~Wilczek for helpful
discussions.
We are grateful to the Department of Energy's
Institute for Nuclear Theory at the University of Washington
for its hospitality and support during the
completion of much of this work.
This work is supported in part  by the U.S. Department
of Energy (D.O.E.) under cooperative research agreement \#DF-FC02-94ER40818.
The work of KR is supported in part by a DOE OJI Award and by the
Alfred P. Sloan Foundation. 
The work of JB is supported by a DOD National Defense Science
and Engineering Graduate Fellowship.
}

\end{document}